\documentclass[11pt,a4paper]{article}
\usepackage{longtable}
\usepackage{amsmath}
\usepackage{amssymb}
\usepackage{graphicx}
\usepackage{bm}
\usepackage{footmisc}
\usepackage{afterpage}
\usepackage{float}
\usepackage{lscape}
\usepackage{longtable}
\usepackage{float}

\newcommand*{\no}{\noindent}
\newcommand*{\bea}{\begin{eqnarray}}
\newcommand*{\eea}{\end{eqnarray}}
\newcommand*{\be}{\begin{equation}}
\newcommand*{\ee}{\end{equation}}
\newcommand*{\pd}{\partial}
\newcommand*{\pdm}{\pd_{\mu}}

\newcommand*{\pref}[1]{(\ref{#1})}

\newcommand*{\nn}{\nonumber}

\newcommand*{\tr}{\mathrm{tr}}

\newcommand{\bma}{\begin{pmatrix}}
\newcommand{\ema}{\end{pmatrix}}

\usepackage[square,comma,sort&compress,numbers]{natbib}

\title{More on the properties of the first Gribov region in Landau gauge}

\author{Axel Maas\\
Institute of Physics, University of Graz,\\
Universit\"atsplatz 5, A-8010 Graz, Austria}

\begin{document}

\maketitle

\begin{abstract}

Complete gauge-fixing beyond perturbation theory in non-Abelian gauge theories is a non-trivial problem. This is particularly evident in covariant gauges, where the Gribov-Singer ambiguity gives an explicit formulation of the problem. In practice, this is a problem if gauge-dependent quantities between different methods, especially lattice and continuum methods, should be compared: Only when treating the Gribov-Singer ambiguity in the same way is the comparison meaningful. To provide a better basis for such a comparison the structure of the first Gribov region in Landau gauge, a subset of all possible gauge copies satisfying the perturbative Landau gauge condition, will be investigated. To this end, lattice gauge theory will be used to investigate a two-dimensional projection of the region for SU(2) Yang-Mills theory in two, three, and four dimensions for a wide range of volumes and discretizations.

\end{abstract}

\section{Introduction}

In gauge theories gauge-fixed correlation functions, like the gauge boson propagator, are excellent tools in intermediate steps of calculations to determine gauge-invariant physics. Thus, gauge-dependent correlation functions have been investigated heavily both in perturbation theory \cite{Bohm:2001yx} and beyond \cite{Alkofer:2000wg,Fischer:2006ub,Binosi:2009qm,Maas:2011se,Boucaud:2011ug,Vandersickel:2012tg}. The key element in the latter calculations has been the judicious combination of lattice gauge theory, functional continuum methods, in particular Dyson-Schwinger equations (DSEs) and functional renormalization group equations (FRGs), effective theories, and perturbation theory \cite{Maas:2011se}. However, this requires to fix a gauge in a controlled way to determine these correlation functions.

Beyond perturbation theory, this becomes complicated. The local gauge conditions employed in perturbation theory are no longer sufficient to uniquely identify a single representative for a gauge orbit. This is the so-called Gribov-Singer ambiguity \cite{Gribov:1977wm,Singer:1978dk,vanBaal:1997gu,vanBaal:1991zw,Dell'Antonio:1991xt}. Here, this ambiguity will be investigated for (SU(2)) Yang-Mills theory in case of the Landau gauge\footnote{Matter fields appear to affect the Gribov-Singer ambiguity \cite{Capri:2012ah,Capri:2013oja,Maas:2010nc,Maas:2011jf}, but this is a different topic.}, i.\ e.\ besides any other gauge conditions all gauge copies also satisfy the perturbative Landau gauge condition.

To resolve this ambiguity, there are in principle two possibilities: Either use non-local gauge conditions to identify a single representative (see e.\ g.\  \cite{Cucchieri:1997dx,Vandersickel:2012tg,Dudal:2014rxa,Dudal:2008sp,Boucaud:2011ug,Schaden:2013ffa,Bogolubsky:2005wf,Bogolubsky:2007bw,Bornyakov:2008yx,Bornyakov:2010nc,Dudal:2009xh,Fachin:1991pu,Fachin:1993qg,Fischer:2008uz,Maas:2009se,Maas:2008ri,Maas:2009ph,Silva:2004bv,Schaden:2014bea,Sternbeck:2012mf,Tissier:2011ey,Zwanziger:1993dh,LlanesEstrada:2012my,Henty:1996kv}) or to average, akin to non-Landau covariant gauges in perturbation theory, over the remaining gauge copies with a suitable weight (see e.\ g.\ \cite{Maas:2011se,Maas:2013vd,Mehta:2009zv,Neuberger:1986xz,vonSmekal:2008en,vonSmekal:2008es,vonSmekal:2007ns,Serreau:2012cg,Parrinello:1990pm,Serreau:2013ila}). Of course, by formally including a $\delta$-function as a weight function, the prior possibility is only a special case of the latter option. Note that the standard minimal Landau gauge appears to be equivalent to an averaging over all Gribov copies inside the first Gribov region with a flat weight function \cite{Maas:2011se,Maas:2013vd}, and thus all treatments without mentioning the residual freedom are implicitly of this type of gauge-fixing. There are numerous of them, especially on the lattice. See \cite{Maas:2011se,Boucaud:2011ug} for reviews.

While such constructions are, at least at the level of an operational definition using algorithms, always possible in lattice gauge theory, this is not the case in the continuum. In fact, the question of how to reconstruct such gauges in continuum calculations, e.\ g.\ in functional methods, has been an important and unresolved question to date \cite{Maas:2011se,Vandersickel:2012tg,Fachin:1991pu,Fischer:2008uz,Maas:2009se,Tissier:2011ey,vonSmekal:2008en,vonSmekal:2008es,vonSmekal:2007ns,Serreau:2012cg,Parrinello:1990pm,Maas:2013vd,Zwanziger:2002ia,Zwanziger:2001kw,Serreau:2013ila,Slavnov:2009mh}. But to be able to compare results on and off the lattice unambiguously, it is necessary to do so. The agreement between both approaches so far \cite{Alkofer:2000wg,Fischer:2006ub,Binosi:2009qm,Maas:2011se,Boucaud:2011ug,Vandersickel:2012tg,Tissier:2011ey,Dudal:2008sp,Serreau:2012cg,Fischer:2008uz,Serreau:2015yna,Huber:2012kd,Eichmann:2014xya,Blum:2014gna,Aguilar:2013xqa,Fister:2013bh} strongly suggests that, at least implicitly, this is already done. It remains to do so explicitly.

One option appears indeed to construct an averaging procedures over Gribov copies, similar to what is done is covariant gauges in perturbation theory. Necessarily, this will require some kind of non-local averaging \cite{Maas:2011se,Parrinello:1990pm,Serreau:2012cg}. In addition, if the resulting term is a surface-term, it is possible to recast it into boundary conditions of functional equations \cite{Guralnik:2007rx}, which indeed appear to discriminate between different solutions of the equations \cite{Maas:2011se,Fischer:2008uz,Boucaud:2008ji}. The question is, whether this can be done in an orbit-independent fashion. For this, it will be necessary to understand the distribution of Gribov copies as a function of the orbit. Contributing towards this goal is the first aim of the present work. To identify suitable averaging procedures it is useful to understand the structure of the first Gribov region. This is the second aim of this work.

The basic structure of the (first) Gribov region is reviewed in section \ref{s:gr}. How it is treated here technically using lattice gauge theory is presented in section \ref{s:tech}. The first result is the number of Gribov copies as a function of dimensions and the lattice parameters in section \ref{s:count}. This is important for the central question of (orbit-independent) normalization. The second question is the structure of the orbit when it comes to possible weight functions. This is presented in section \ref{s:struct}. Results and consequences are then wrapped up in section \ref{s:sum}.

Similar investigations on the structure of the Gribov region have also been performed in \cite{Cucchieri:1997dx,Hughes:2012hg,Greensite:2010hn,Greensite:2004ur,Mehta:2011sp,Mehta:2014jla,Sternbeck:2015gia,Sternbeck:2005vs,Cucchieri:2013nja,Cucchieri:2013xka,Sternbeck:2012mf}, though all having a different focus and, partly approach, as in the present work, and are therefore complementary. Some preliminary results on the topic of this work can be found in \cite{Maas:2011se,Maas:2011ba,Maas:2009se,Maas:2010wb}. Using the information gained in this work to construct various completions of the Landau gauge along the lines of \cite{Maas:2009se,Maas:2013vd,Parrinello:1990pm,Serreau:2012cg} will be presented in an upcoming work \cite{Maas:unpublished}.

\section{The (first) Gribov region}\label{s:gr}

The existence of the first Gribov region in Landau gauge is arguably the geometrically most remarkable feature of (Euclidean) Yang-Mills theory. The first Gribov region in Landau gauge is defined \cite{Gribov:1977wm} as the set of all gauge copies on a given gauge orbit which satisfy the perturbative Landau gauge condition $\pdm A_\mu^a$ and for which the Faddeev-Popov operator
\be
M^{ab}=-\pdm D_\mu^{ab}=-\pd^2\delta^{ab}+gf^{abc} A_\mu^c=-D_\mu^{ab}\pdm\nn
\ee
\no is positive semi-definite, i.\ e.\ it has only zero or positive eigenvalues. There is at least one gauge copy on every gauge orbit which satisfies both conditions \cite{Dell'Antonio:1991xt}, and thus a restriction to the first Gribov region is meaningful. Furthermore, and this is the most distinct characteristic, this region is bounded \cite{Zwanziger:1982na}, and thus only bounded field configurations are required to describe all of the physics of Yang-Mills theory. The boundary can be shown to be not a smooth manifold, but exhibits, e.\ g.\ cone-like singularities and cusps \cite{Greensite:2004ur}. The Faddeev-Popov operator is also the inverse ghost propagator, which is therefore linked to the eigenspectrum. In fact, there is evidence that the ghost propagator at the lowest momentum accessible on a finite lattice \cite{Cucchieri:2006tf} is essentially dominated by the lowest eigenvalue, at least for the largest volumes \cite{Sternbeck:2012mf,Cucchieri:2008fc,Cucchieri:2013nja}, though this is not true at small volumes \cite{Sternbeck:2005vs}. Therefore, the ghost propagator at the lowest momentum can be taken as an approximate and indirect measure for the behavior of the lowest eigenvalue of the Faddeev-Popov operator \cite{Maas:2009se}.

This restriction is not giving a unique gauge copy for every gauge orbit, and gauge orbits can have multiple gauge copies inside the first Gribov region \cite{Semenov:1982,vanBaal:1997gu}. These are called Gribov copies, though technically they are just a subset of ordinary gauge copies, and physically indistinct from all remaining gauge copies outside the first Gribov region. Still, here Gribov copies will denote the gauge copies within or on the boundary of the first Gribov region.

The Gribov copies are non-perturbative, in the sense that one Gribov copy cannot be reached from another one by an infinitesimal gauge transformation \cite{Gribov:1977wm}. However, every Gribov copy inside the first Gribov region can be reached from any other Gribov copy by a gauge transformation. Thus, to every gauge orbit, there is a restricted set of gauge transformation which changes between the Gribov copies. Since the Gribov copies are physically indistinct, this defines a residual gauge symmetry inside the first Gribov region: The set of Gribov copies for each gauge orbit together with the set of gauge transformations transforming between them. Conceptually \cite{Maas:2013vd,Maas:2012ct}, this is the same as the BRST symmetry of ordinary perturbation theory \cite{Bohm:2001yx}, which is also only a set of gauge transformations which leaves the gauge condition intact. However, in the present case, an explicit definition of the corresponding transformation is still lacking, and will likely be a non-local one \cite{Sorella:2009vt}.

There are other Gribov regions, which surround the first Gribov region. They are defined by the number of negative eigenvalues of the Faddeev-Popov operator, which develops another one when passing a Gribov horizon by letting the lowest positive one pass through zero to become negative. These Gribov regions are fully embedded into each other. They are not subject of this work, though it is hypothesized that they may play a role in recovering the conventional BRST symmetry beyond perturbation theory \cite{Maas:2012ct,vonSmekal:2008en,vonSmekal:2008es,Fischer:2008uz}.

The first Gribov region can also be characterized by the condition that every Gribov copy minimizes the functional
\be
F[A_\mu]=-\int d^dx A_\mu^a A_\mu^a=\sum_{x,\mu}\Re\tr U_\mu(x)\label{f},
\ee
\no where $A$ is the gauge field and $U$ the link variable. This quantity has as first derivative the Landau gauge condition and as Hessian the Faddeev-Popov operator. The expectation value of this quantity is just the momentum-integrated gluon propagator \cite{Maas:2008ri}, and thus the first Gribov region is the region of minimal (integrated) Gluon propagators. However, this quantity is a composite operator and divergent in the continuum limit, and thus requires renormalization.

Since these are minima, there is the possibility of a unique absolute minimum. Such an absolute minimum exists, and the set of all Gribov copies which are absolute minima is called the fundamental modular region (FMR) \cite{Zwanziger:1993dh}. In principle, it is possible that the minima become degenerate in the thermodynamic limit, as happens, e.\ g.\ in U(1) gauge theory \cite{deForcrand:1994mz}.

In fact, since there is at least one minimum, and therefore always at least one absolute minimum, on every gauge orbit, every gauge orbit passes through this region. Moreover, it can be shown that this region is, as the first Gribov region, bounded and convex \cite{Zwanziger:1993dh}. Furthermore, it contains the origin, and therefore is centered inside the first Gribov region. On any finite volume, its boundary does not touch the one of the first Gribov region, i.\ e.\ the Faddeev-Popov operator has no zero eigenvalue inside it in a finite volume. In the infinite-volume limit, this region has a common boundary with the first Gribov region \cite{Zwanziger:1993dh}, though this common boundary is a subset of the boundary of the first Gribov region, the so-called Gribov horizon. Note that the boundary of the fundamental modular region is rather complex, as on its boundary degeneracies may arise, which have to be taken care of \cite{vanBaal:1997gu}.

\section{Technical setup}\label{s:tech}

The technical setup for this investigation is essentially as described in \cite{Cucchieri:2006tf,Maas:2009se}. The simulations are performed for the Wilson action of $d=2,3,4$-dimensional Yang-Mills theories on a lattice of size $N^d$ for some bare gauge coupling $\beta$, using a mixture of overrelaxation and heat-bath sweeps \cite{Cucchieri:2006tf}. The lattice spacing is set by assigning the string tension a value of $(440$ MeV$)^2$, as described in \cite{Cucchieri:2006tf,Maas:2008ri}.

\begin{longtable}{|c|c|c|c|c|c|c|c|}
\caption{\label{tcgf}Number and parameters of the configurations used, ordered by dimension, lattice spacing, and physical volume. $p_\text{min}$ is the smallest momentum on the given lattice, and thus the one at which the ghost propagator has been evaluated. In all cases $2(10N+100(d-1))$ thermalization sweeps and $2(N+10(d-1))$ decorrelation sweeps of mixed updates \cite{Cucchieri:2006tf} have been performed, and auto-correlation times of local observables have been monitored to be at or below one sweep. The number of configurations were selected such as to have a reasonable small statistical error for the ghost propagator at the lowest momentum. The number $N_r$ was chosen such that, given the results from lattices with smaller physical volumes and/or coarser discretizations, that the total fraction of identified genuine Gribov copies should be substantial. The number of configurations had to be also chosen large enough such that so-called exceptional gauge orbits with particular extreme Gribov copies, i.\ e.\ with very large coordinates, were (marginally) sufficiently sampled as well. In total ${\cal O}(10^5)$ configurations have been obtained and ${\cal O}(10^7)$ gauge-fixings have been performed.}\\
\hline
$d$	& $N$	& $\beta$	& $a^{-1}$ [MeV]	& L [fm]	& $p_\text{min}$ [MeV]	& $N_r$	& config.	\endfirsthead
\hline
\multicolumn{8}{|l|}{Table \ref{tcgf} continued}\\
\hline
$d$	& $N$	& $\beta$	& $a^{-1}$ [MeV]	& L [fm]	& $p_\text{min}$ [MeV]	& $N_r$	& config.	\endhead
\hline
\multicolumn{8}{|r|}{Continued on next page}\\
\hline\endfoot
\endlastfoot
\hline
2	& 92	& 6.23		& 863			& 21		& 58.9			& 21	& 2761		\cr
\hline
2	& 80	& 6.40		& 875			& 18		& 68.7			& 20	& 2634		\cr
\hline
2	& 58	& 6.45		& 879			& 13		& 95.2			& 20	& 2220		\cr
\hline
2	& 18	& 6.55		& 886			& 4.0		& 299			& 20	& 1720		\cr
\hline
2	& 34	& 6.64		& 893			& 7.5		& 165			& 20	& 1590		\cr
\hline
2	& 68	& 6.64		& 893			& 15		& 82.5			& 20	& 2503		\cr
\hline
2	& 10	& 6.68		& 895			& 2.2		& 553			& 20	& 2234		\cr
\hline
2	& 50	& 6.68		& 895			& 11		& 112			& 20	& 2107		\cr
\hline
2	& 26	& 6.72		& 898			& 5.7		& 216			& 20	& 1320		\cr
\hline
2	& 42	& 6.73		& 899			& 9.2		& 134			& 20	& 1841		\cr
\hline
2	& 106	& 8.13		& 994			& 21		& 58.9			& 22	& 2478		\cr
\hline
2	& 92	& 8.33		& 1010			& 18		& 69.0			& 21	& 5166		\cr
\hline
2	& 68	& 8.70		& 1030			& 13		& 95.2			& 20	& 4624		\cr
\hline
2	& 58	& 8.83		& 1040			& 11		& 113			& 20	& 2220		\cr
\hline
2	& 80	& 9.03		& 1050			& 15		& 82.4			& 20	& 2388		\cr
\hline
2	& 50	& 9.36		& 1070			& 9.2		& 134			& 20	& 2107		\cr
\hline
2	& 42	& 9.91		& 1100			& 7.5		& 164			& 20	& 1841		\cr
\hline
2	& 122	& 10.6		& 1140			& 21		& 58.7			& 23	& 2806		\cr
\hline
2	& 106	& 10.9		& 1160			& 18		& 68.7			& 22	& 2478		\cr
\hline
2	& 34	& 11.1		& 1170			& 5.7		& 216			& 20	& 1590		\cr
\hline
2	& 92	& 11.7		& 1200			& 15		& 81.9			& 21	& 2489		\cr
\hline
2	& 80	& 11.8		& 1210			& 13		& 95.0			& 20	& 4865		\cr
\hline
2	& 68	& 11.9		& 1210			& 11		& 112			& 20	& 2780		\cr
\hline
2	& 58	& 12.4		& 1240			& 9.2		& 134			& 20	& 2331		\cr
\hline
2	& 26	& 13.1		& 1280			& 4.0		& 309			& 20	& 1320		\cr
\hline
2	& 50	& 13.8		& 1310			& 5.7		& 165			& 20	& 2107		\cr
\hline
2	& 122	& 14.3		& 1340			& 18		& 69.0			& 23	& 1536		\cr
\hline
2	& 92	& 15.5		& 1390			& 13		& 94.9			& 21	& 2522		\cr
\hline
2	& 106	& 15.5		& 1390			& 15		& 82.4			& 22	& 2478		\cr
\hline
2	& 80	& 16.3		& 1430			& 11		& 112			& 20	& 4980		\cr
\hline
2	& 42	& 16.8		& 1450			& 5.7		& 217			& 20	& 1829		\cr
\hline
2	& 68	& 16.9		& 1460			& 9.2		& 135			& 20	& 2385		\cr
\hline
2	& 58	& 18.4		& 1520			& 7.5		& 165			& 20	& 8929		\cr
\hline
2	& 122	& 20.3		& 1600			& 15		& 82.4			& 23	& 2833		\cr
\hline
2	& 106	& 20.4		& 1600			& 13		& 94.9			& 22	& 2814		\cr
\hline
2	& 18	& 20.6		& 1610			& 2.2		& 559			& 20	& 1720		\cr
\hline
2	& 92	& 21.5		& 1650			& 11		& 113			& 21	& 2583		\cr
\hline
2	& 34	& 22.2		& 1670			& 4.0		& 308			& 20	& 1590		\cr
\hline
2	& 80	& 23.2		& 1710			& 9.2		& 134			& 20	& 2370		\cr
\hline
2	& 50	& 23.6		& 1730			& 5.7		& 217			& 20	& 2103		\cr
\hline
2	& 68	& 25.2		& 1790			& 7.5		& 165			& 20	& 2660		\cr
\hline
2	& 122	& 26.9		& 1850			& 13		& 95.3			& 23	& 1625		\cr
\hline
2	& 106	& 28.4		& 1900			& 11		& 113			& 22	& 2508		\cr
\hline
2	& 92	& 30.5		& 1970			& 9.2		& 135			& 21	& 2694		\cr
\hline
2	& 58	& 31.6		& 2000			& 5.7		& 217			& 20	& 2260		\cr
\hline
2	& 42	& 33.6		& 2070			& 4.0		& 309			& 20	& 1814		\cr
\hline
2	& 80	& 34.7		& 2100			& 7.5		& 165			& 20	& 2466		\cr
\hline
2	& 122	& 37.4		& 2180			& 11		& 112			& 23	& 2807		\cr
\hline
2	& 106	& 40.4		& 2270			& 9.2		& 135			& 22	& 2604		\cr
\hline
2	& 68	& 43.2		& 2350			& 5.7		& 217			& 20	& 2712		\cr
\hline
2	& 92	& 45.7		& 2420			& 7.5		& 165			& 21	& 2520		\cr
\hline
2	& 26	& 46.5		& 2440			& 2.2		& 588			& 20	& 1320		\cr
\hline
2	& 50	& 47.4		& 2460			& 4.0		& 309			& 20	& 2102		\cr
\hline
2	& 122	& 53.3		& 2610			& 9.2		& 134			& 23	& 2592		\cr
\hline
2	& 80	& 59.7		& 2760			& 5.7		& 217			& 20	& 2650		\cr
\hline
2	& 106	& 60.5		& 2780			& 7.5		& 165			& 22	& 2546		\cr
\hline
2	& 58	& 63.7		& 2860			& 4.0		& 310			& 20	& 3224		\cr
\hline
2	& 34	& 72.3		& 3040			& 2.2		& 561			& 20	& 1590		\cr
\hline
2	& 92	& 78.8		& 3180			& 5.7		& 217			& 21	& 2650		\cr
\hline
2	& 122	& 80		& 3200			& 7.5		& 165			& 23	& 2925		\cr
\hline
2	& 68	& 87.3		& 3350			& 4.0		& 309			& 20	& 2790		\cr
\hline
2	& 106	& 104		& 3650			& 5.7		& 216			& 22	& 2574		\cr
\hline
2	& 42	& 110		& 3760			& 2.2		& 562			& 20	& 1787		\cr
\hline
2	& 80	& 120		& 3930			& 4.0		& 309			& 20	& 2656		\cr
\hline
2	& 122	& 138		& 4210			& 5.7		& 217			& 22	& 2833		\cr
\hline
2	& 50	& 155		& 4470			& 2.2		& 561			& 20	& 2222		\cr
\hline
2	& 92	& 159		& 4520			& 4.0		& 309			& 21	& 2606		\cr
\hline
2	& 58	& 209		& 5190			& 2.2		& 562			& 20	& 3236		\cr
\hline
2	& 106	& 211		& 5210			& 4.0		& 309			& 22	& 2679		\cr
\hline
2	& 122 	& 280		& 6010			& 4.0		& 310			& 23	& 2624		\cr
\hline
2	& 68	& 287		& 6090			& 2.2		& 563			& 20	& 2940		\cr
\hline
2	& 80	& 398		& 7160			& 2.2		& 562			& 20	& 2646		\cr
\hline
2	& 92	& 526		& 8240			& 2.2		& 563			& 21	& 2557		\cr
\hline
2	& 106	& 698		& 9490			& 2.2		& 562			& 22	& 2520		\cr
\hline
2	& 122	& 925		& 10900			& 2.2		& 561			& 23	& 2623		\cr
\hline
\hline
3	& 8	& 3.40		& 874			& 1.8		& 669			& 20	& 2110		\cr
\hline
3	& 14	& 3.44		& 887			& 3.1		& 395			& 20	& 1650		\cr
\hline
3	& 20	& 3.46		& 894			& 4.4		& 280			& 22	& 1400		\cr
\hline
3	& 26	& 3.47		& 897			& 5.7		& 216			& 39	& 3795		\cr
\hline
3	& 36	& 3.47		& 897			& 7.9		& 156			& 66	& 1688		\cr
\hline
3	& 32	& 3.48		& 900			& 7.0		& 176			& 59	& 1627		\cr
\hline
3	& 36	& 3.82		& 1010			& 7.0		& 176			& 66	& 1491		\cr
\hline
3	& 42	& 3.92		& 1070			& 7.9		& 160			& 72	& 1583		\cr
\hline
3	& 32	& 4.10		& 1100			& 5.7		& 216			& 59	& 2791		\cr
\hline
3	& 42	& 4.33		& 1180			& 7.0		& 176			& 71	& 1447		\cr
\hline
3	& 26	& 4.28		& 1160			& 4.4		& 280			& 39	& 1291		\cr
\hline
3	& 36	& 4.52		& 1240			& 5.7		& 216			& 66	& 1496		\cr
\hline
3	& 20	& 4.60		& 1270			& 3.1		& 397			& 22	& 1380		\cr
\hline
3	& 32	& 5.09		& 1430			& 4.4		& 280			& 57	& 2744		\cr
\hline
3	& 42	& 5.15		& 1450			& 5.7		& 217			& 71	& 1621		\cr
\hline
3	& 14	& 5.39		& 1530			& 1.8		& 680			& 20	& 1720		\cr
\hline
3	& 36	& 5.64		& 1610			& 4.4		& 281			& 63	& 1633		\cr
\hline
3	& 26	& 5.76		& 1650			& 3.1		& 398			& 30	& 1334		\cr
\hline
3	& 42	& 6.45		& 1880			& 4.4		& 281			& 64	& 1707		\cr
\hline
3	& 32	& 6.91		& 2030			& 3.1		& 398			& 40	& 1585		\cr
\hline
3	& 48	& 7.27		& 2150			& 4.4		& 281			& 70	& 1535		\cr
\hline
3	& 20	& 7.39		& 2190			& 1.8		& 685			& 20	& 1450		\cr
\hline
3	& 36	& 7.69		& 2290			& 3.1		& 399			& 45	& 1478		\cr
\hline
3	& 42	& 8.84		& 2670			& 3.1		& 399			& 46	& 1592		\cr
\hline
3	& 26	& 9.38		& 2840			& 1.8		& 685			& 20	& 1315		\cr
\hline
3	& 48	& 10.0		& 3050			& 3.1		& 399			& 47	& 2037		\cr
\hline
3	& 32	& 11.3		& 3480			& 1.8		& 682			& 20	& 1417		\cr
\hline
3	& 36	& 12.7		& 3940			& 1.8		& 687			& 20	& 1370		\cr
\hline
3	& 42	& 14.6		& 4570			& 1.8		& 683			& 20	& 1699		\cr
\hline
3	& 48	& 16.6		& 5220			& 1.8		& 683			& 20	& 1877		\cr
\hline
3	& 54	& 18.6		& 5880			& 1.8		& 682			& 21	& 1832		\cr
\hline
3	& 60	& 20.6		& 6540			& 1.8		& 685			& 22	& 2044		\cr
\hline
\hline
4	& 14	& 2.179		& 889			& 3.1		& 396			& 27	& 1082		\cr
\hline
4	& 10	& 2.181		& 894			& 2.2		& 553			& 20	& 1450		\cr
\hline
4	& 6	& 2.188		& 908			& 1.3		& 908			& 20	& 1620		\cr
\hline
4	& 18	& 2.188		& 908			& 3.9		& 315			& 54	& 1615		\cr
\hline
4	& 18	& 2.279		& 1140			& 3.1		& 396			& 54	& 2248		\cr
\hline
4	& 14	& 2.311		& 1250			& 2.2		& 556			& 27	& 1033		\cr
\hline
4	& 22	& 2.349		& 1400			& 3.1		& 398			& 76	& 1488		\cr
\hline
4	& 10	& 2.376		& 1520			& 1.3		& 939			& 20	& 1450		\cr
\hline
4	& 18	& 2.395		& 1610			& 2.2		& 559			& 40	& 1258		\cr
\hline
4	& 22	& 2.457		& 1960			& 2.2		& 558			& 50	& 1242		\cr
\hline
4	& 14	& 2.480		& 2120			& 1.3		& 943			& 20	& 1225		\cr
\hline
4	& 18	& 2.552		& 2720			& 1.3		& 945			& 20	& 1175		\cr
\hline
4	& 22	& 2.609		& 3330			& 1.3		& 948			& 20	& 1355		\cr
\hline
4	& 26	& 2.656		& 3930			& 1.3		& 947			& 21	& 1335		\cr
\hline
\end{longtable}

The resulting set of configurations is then gauge fixed using a self-adapting stochastic overrelaxation algorithm \cite{Cucchieri:2006tf}. To investigate different Gribov copies, this is performed using $N_r$ restarts with random seeds \cite{Cucchieri:1997dx}. This operation is rather expensive, and thus only a limited set of configurations and lattice parameters with different $N_r$ could be used to track the development. These parameters are listed in table \ref{tcgf}. Note that in the following very often only particular sets of parameters are shown as an examples, but all sets of parameters have been investigated  and showed the same behavior. The so created $N_r$ gauge-fixed gauge copies of any configuration are not necessarily distinct, which will be addressed below in section \ref{s:count}.

Nonetheless, for every so created gauge copy the quantities \pref{f}, obtained from the links, and the ghost propagator at the lowest momentum, using the method described in \cite{Cucchieri:2006tf}, have been calculated. Note that, in contrast to \cite{Maas:2009se} the ghost propagator has been determined using a plane-wave source \cite{Cucchieri:2006tf}, to reduce statistical fluctuations, and as the average in color space and over all possible momentum representations, i.\ e.\ the edge momenta with only one non-vanishing component. This reduces the consequences of the violation of rotational symmetry on small lattices. In the course of this investigation, it has been found that on the largest two-dimensional lattices the result, within statistical error, was the same whether this average over momentum-representatives was done or not. This is expected as that with lower and lower momenta the difference should diminish as the ghost propagator on every configuration should have a unique value at zero momentum, though at finite momentum this needs not be the case. Even on the largest volumes in 3 and 4 dimensions, this was not yet reached, though the difference remains minor, a few times the statistical error, of the order of a few percent. Therefore results without averaging, as was done in \cite{Maas:2009se}, would have no qualitative, and almost no quantitative impact, for the results presented here. Nonetheless, all results here have been obtained with momentum-averaging.

These two quantities are kept to characterize the gauge copy. Of course, it would be desirable to rather keep the full configurations: Two Gribov copies are only really distinct, if there is at least a single point where the field configurations differ, up to permitted translations, rotations, and global color rotations. In practice, it is impossible to check for all these transformations. Also, the required disk space is simply not available for the amount of lattice parameters and statistics employed here. However, it appears that two different Gribov copies differ appreciably in a finite region rather than isolated points \cite{Heinzl:2007cp}. Therefore there should also be a finite difference in many quantities. This should be therefore sufficient to distinguish two copies. This will be discussed more next.

\section{Gribov copies}\label{s:count}

\subsection{Distinguishing Gribov copies}

The most basic question is how many Gribov copies there are \cite{Mehta:2009zv,Mehta:2011sp,Hughes:2012hg,Maas:2009se,Maas:2011se}. However, as noted, to distinguish two Gribov copies from each other in practical calculations is a non-trivial problem. In principle, it would be necessary to check at each space-time point for differences, within the numerical accuracy, taking all possible global transformations into account. Already the memory limitations are prohibitive in practice for any appreciable number of Gribov copies.

Here, another approach is used. In principle, two Gribov copies will in general differ also in at least one gauge-dependent correlation function, which can be considered as the moments of the configuration. Thus, knowledge of the gauge-dependent correlation functions can also be used to characterize a Gribov copy. But, again, in practice only a finite number of them can be calculated. And if only a finite number is compared then it is by far not clear whether two distinct Gribov copies will also be distinct in this limited set. In practice, so far almost always only one function at a single momentum value, mostly \pref{f}, had been used. Here, following up on the preliminary investigations \cite{Maas:2009se,Maas:2011se}, for the first time a systematic investigation will be performed using two quantities simultaneously,
\bea
F&=&-\frac{1}{V}\int d^d x A_\mu^a A_\mu^a\nn=1-\frac{1}{N_c}\sum_{x,\mu}\tr U_\mu(x)\\
b&=&\tilde{Z}_3\frac{1}{dN_c}\sum_i G^{aa}(p(i)_\text{min}^2)\nn,
\eea
\no where $G$ is the ghost dressing function \cite{Cucchieri:2006tf,Maas:2011se}, and the renormalization constant $\tilde{Z}_3$ is only different from one in four dimensions, and will usually drop out below, since mostly ratios will be considered\footnote{Note that this implicitly assumes that renormalization constants do not differ depending on the selection of Gribov copies. At least within achievable statistical errors, this should be the case if the renormalization is performed at sufficiently large momenta where the influence of Gribov copies is negligible for the ghost propagator \cite{Maas:2011se}.}. If it used explicitly, it is obtained by requiring $\tilde{Z}_3G((2$ GeV$)^2)=1$. The sum is over all possible directions for the minimal edge momentum. Effectively, this is a projection of the Gribov region to a two-dimensional space with coordinates $F$ and $b$. In principle also $F$, as a composite operator, requires renormalization \cite{Itzykson:1980rh}.

In the following, two Gribov copies will be considered to be distinct if their difference in $F$ or $b$ exceeds a certain threshold $\epsilon$. Hence, some Gribov copies, which are different, will not be recognized as such. Thus, all results which indicate a sensitivity to the choice of Gribov copies yields only a lower limit to this sensitivity, even if all Gribov copies of any given residual gauge orbit would be determined. So, the value of $\epsilon$ has to be set carefully. For a properly renormalized quantity, $\epsilon$ could be the numerical accuracy, which is usually significantly less than the machine precision, giving possible cancellations in the sums and imprecisions in the gauge-fixing process. 

However, the values of both $F$ and $b$ drift due to renormalization as well as discretization and finite-volume effects. Thus, at fixed numerical accuracy two originally distinct Gribov copies could appear equal if the lattice settings are changed, because their difference will no longer be resolvable.

To study this behavior, a useful option is to determine for all $N_r$ Gribov copies of a given gauge orbit their distance with respect to each other. For any of the two distinction parameters $D=F,b$ it is therefore useful to investigate the relative difference of their values for two Gribov copies $i$ and $j$ on the same orbit
\be
\Delta D=\frac{|D_i-D_j|}{|\langle\langle D\rangle\rangle|}\nn,
\ee
\no where $\langle\langle D\rangle\rangle$ is for now the both orbit and copies averaged value. Note that both distinction parameters are global color invariants and are also isotropically lattice-averaged. Any global color rotation, translation or rotation therefore does not change their values, even on a single configuration. Taking the ratio removes any kind of trivial kinematic factors, lattice spacing factors, and, provided the renormalization is performed at sufficiently large momenta, multiplicative renormalization factors.

\begin{figure}
\includegraphics[width=\linewidth]{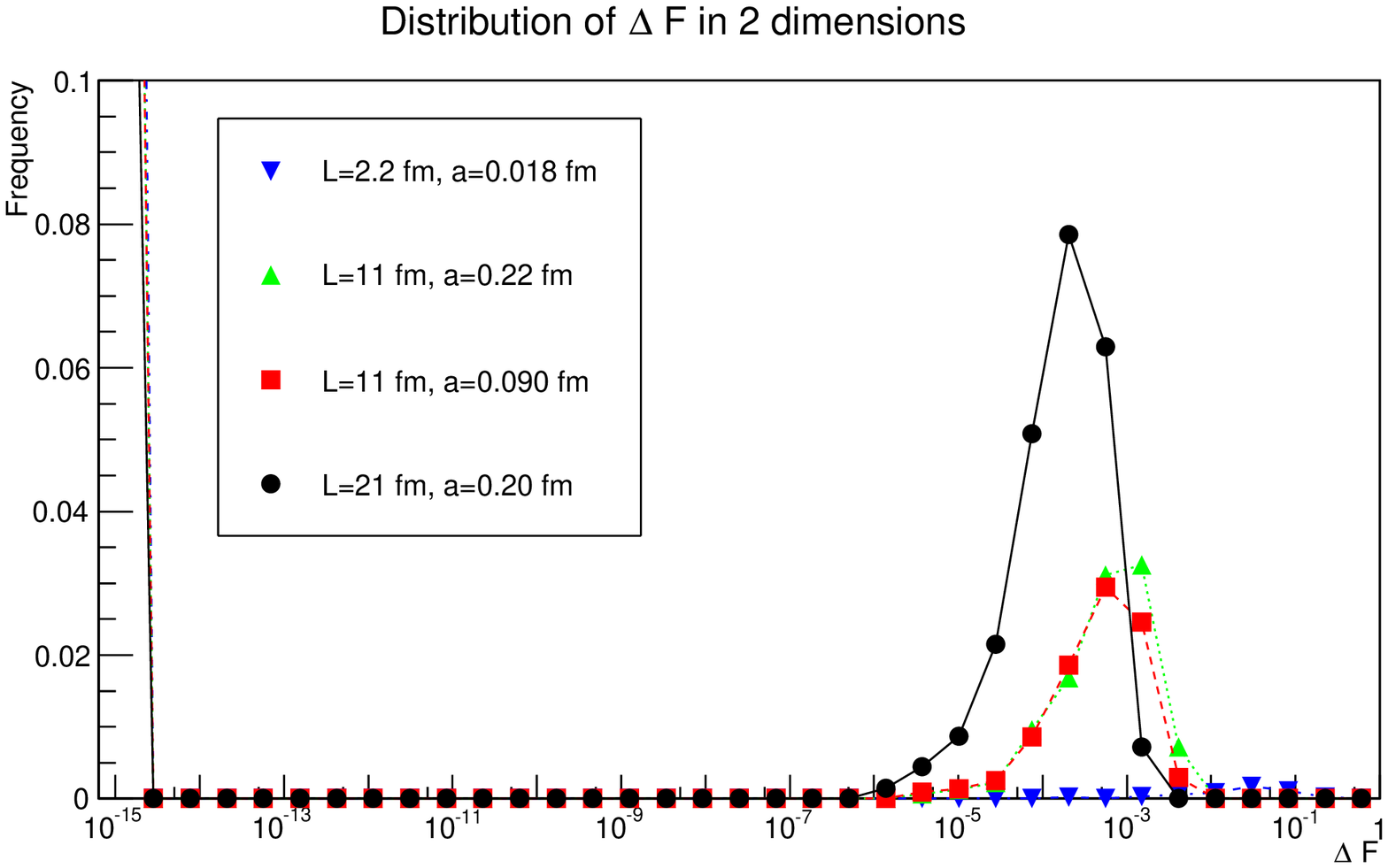}\\
\includegraphics[width=\linewidth]{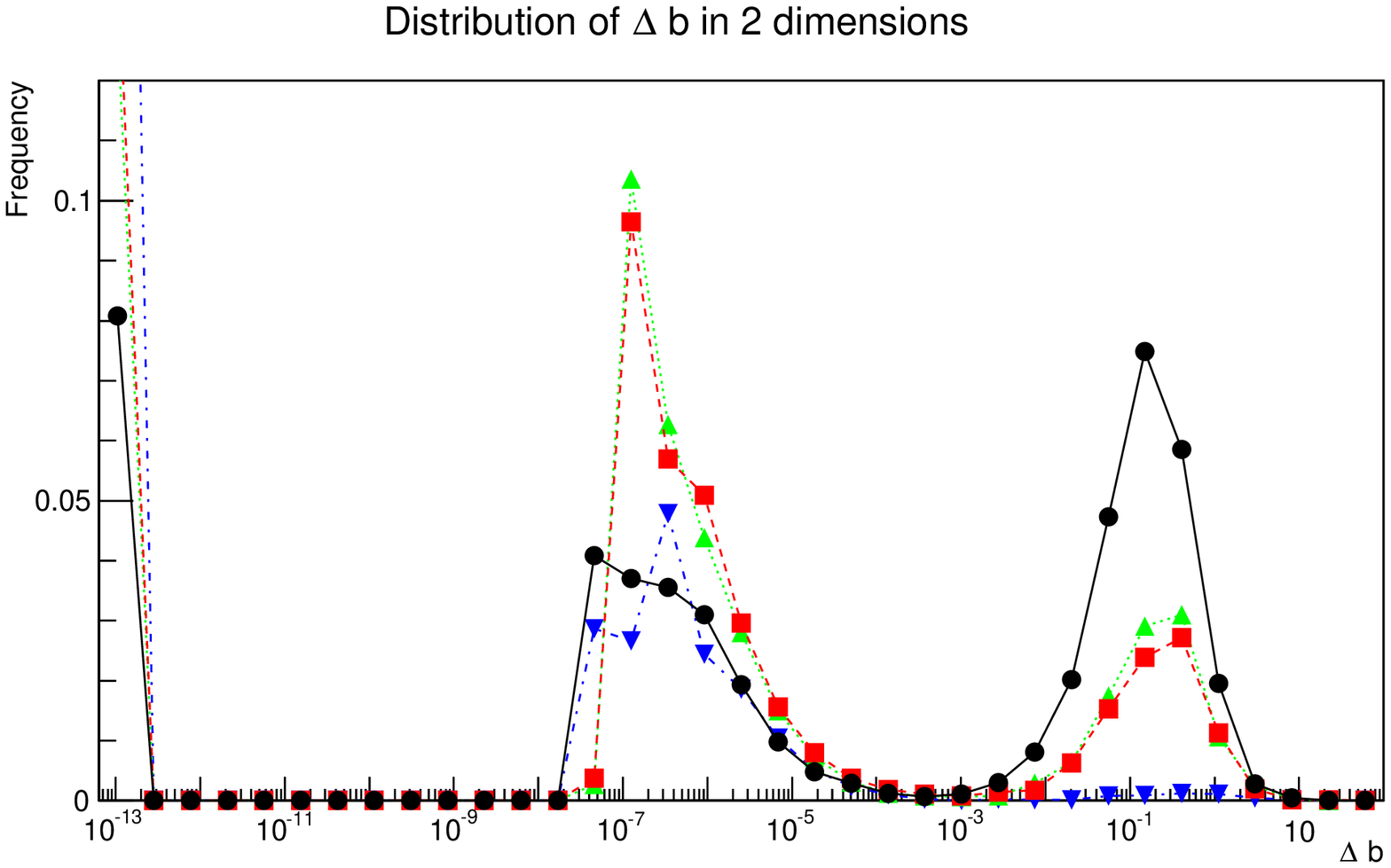}
\caption{\label{fig:2ddiff}Difference of distinction parameter, either $F$ (top panel) or $b$ (bottom panel) for different Gribov copies on the same orbit in two dimensions, for various lattice spacings and volumes, normalized to the orbit and Gribov copy averaged value.}
\end{figure}

\begin{figure}
\includegraphics[width=\linewidth]{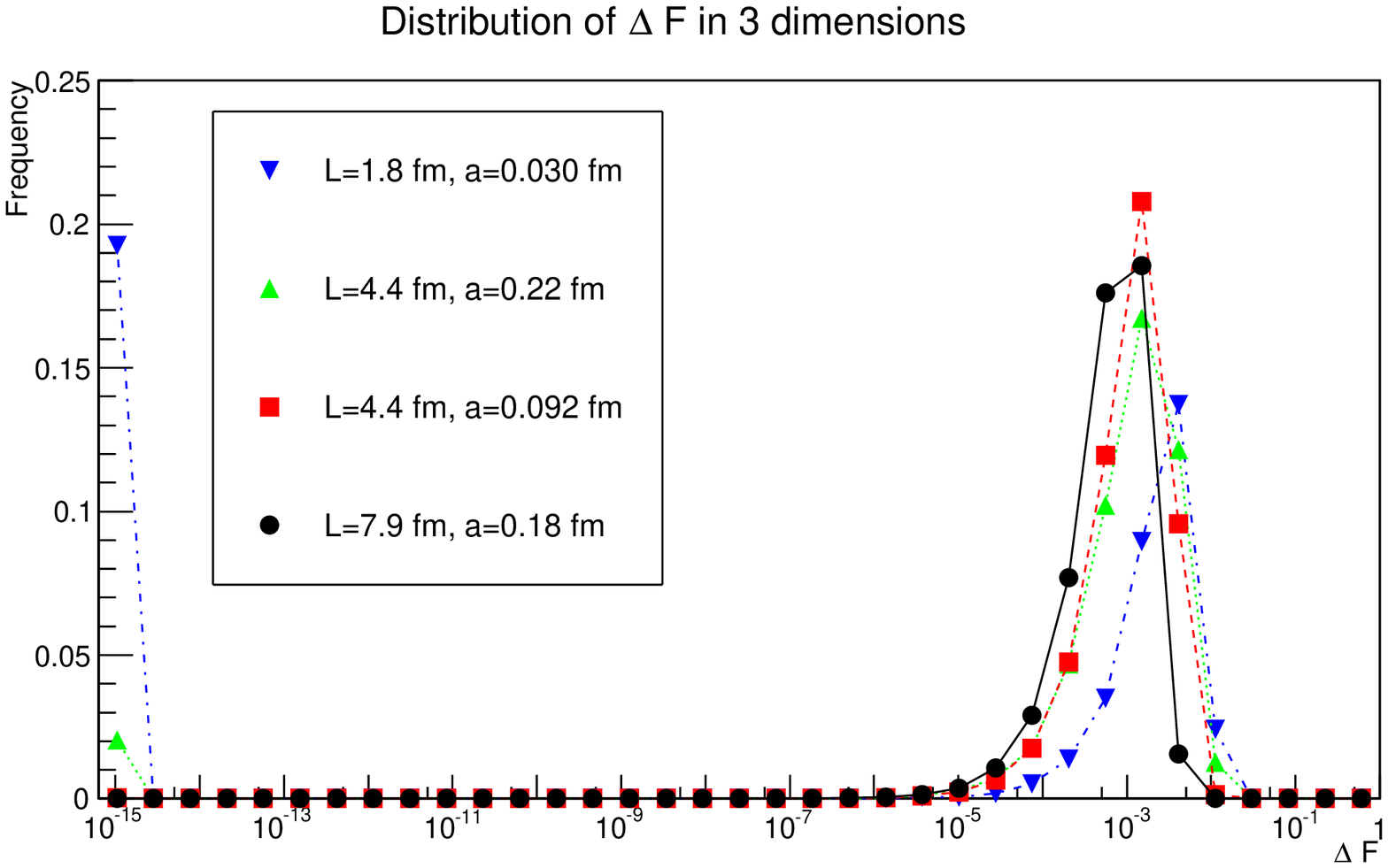}\\
\includegraphics[width=\linewidth]{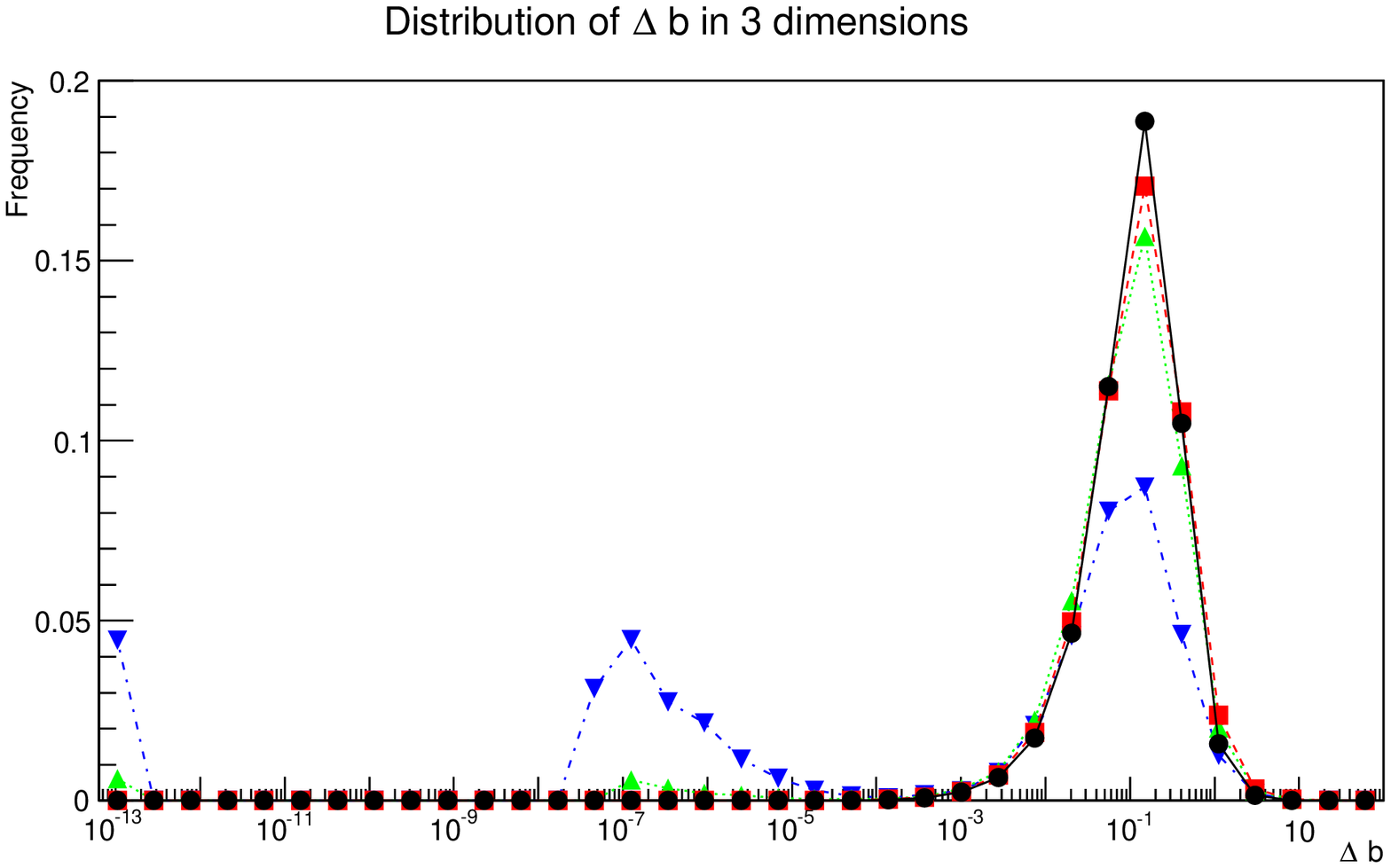}
\caption{\label{fig:3ddiff}Difference of distinction parameter, either $F$ (top panel) or $b$ (bottom panel) for different Gribov copies on the same orbit in three dimensions, for various lattice spacings and volumes, normalized to the orbit and Gribov copy averaged value.}
\end{figure}

\begin{figure}
\includegraphics[width=\linewidth]{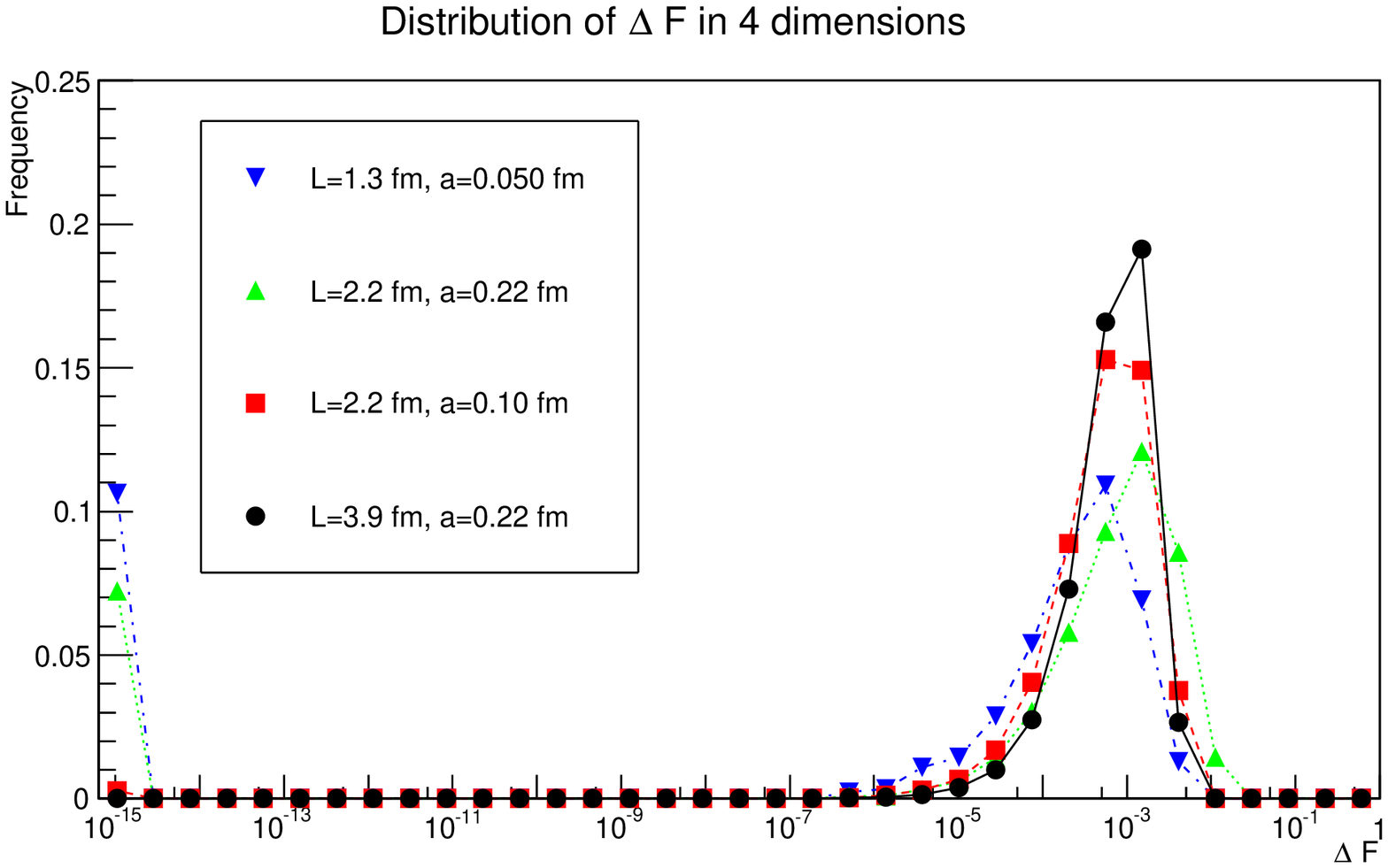}\\
\includegraphics[width=\linewidth]{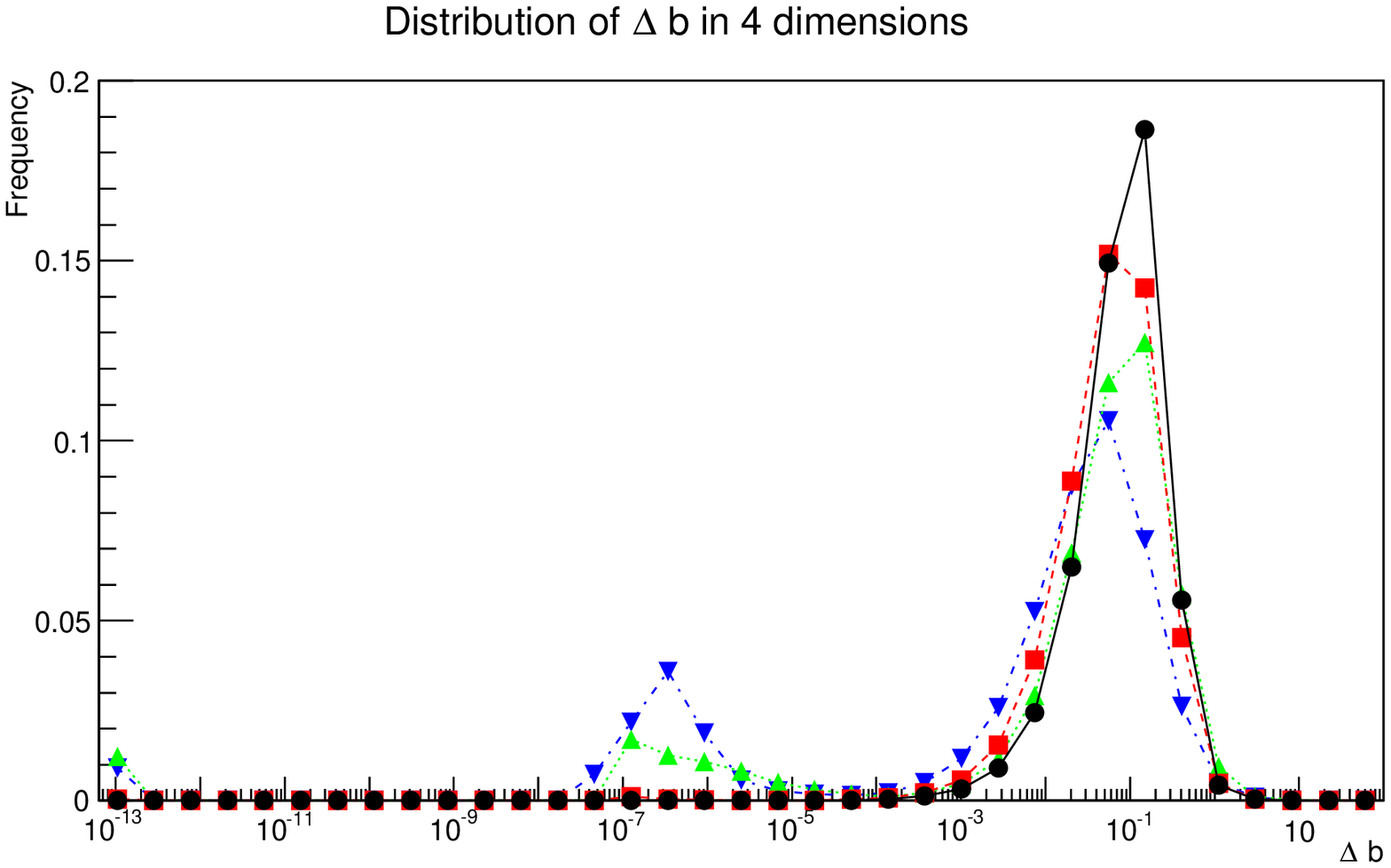}
\caption{\label{fig:4ddiff}Difference of distinction parameter, either $F$ (top panel) or $b$ (bottom panel) for different Gribov copies on the same orbit in four dimensions, for various lattice spacings and volumes, normalized to the orbit and Gribov copy averaged value.}
\end{figure}

The results are shown in figures \ref{fig:2ddiff}-\ref{fig:4ddiff}. For this purpose, the results are binned in $e$-folds. The first observation is a large gap between a structure and the bin of lowest value. In fact, for all elements of the lowest bin the difference is zero within numerical accuracy. Thus, for all purposes these Gribov copies have the same value of their respective distinction parameters. Given the large gaps involving many orders of magnitude, where the bins are in fact unpopulated, it can be safely assumed that the equality is not just numerical coincidence, but indeed true, and thus these are the same Gribov copies. The remaining Gribov copies can be considered to be distinct.

Interestingly, there are Gribov copies which are only indistinct with respect to one parameter, but not with respect to the other\footnote{This was not seen in \cite{Maas:2009se}, and erroneously assumed not to be the case. The reason is that a much lower cut has been used to define two Gribov copies as equal there. Thus, the present, much more precise, results supersede those of \cite{Maas:2009se}}. Predominantly, the copies are indistinct in $F$, but different in $b$, depending on the lattice parameters about 10$^2$-10$^3$ more frequently. This shows that $b$ is a finer distinction of Gribov copies than $F$.

Concerning the distribution for the true Gribov copies, the differences in $\Delta F$ show some dependency on volume and discretization, especially in two dimensions, and a somewhat asymmetric distribution, centered at about $10^{-3}-10^{-4}$ for large volumes. The difference seems to shrink with increasing volume and discretization, i.\ e.\ the distinction between two Gribov copies becomes less. This is to be expected if Gribov copies indeed tend to differ only over some finite patch of space-time \cite{Heinzl:2007cp}.

The situation for the distinction parameter $b$ is interesting. For small volumes, it shows a double-peak structure, which becomes less pronounced for larger volumes, and at fixed volumes for finer discretization. The second peak is at substantially smaller differences. It appears reasonable that they may arise by so-called lattice Gribov copies \cite{Giusti:2001xf}, i.\ e.\ artificial Gribov copies introduced due to the lattice approximation itself, which vanish in the thermodynamic limit. At any rate, they are a small effect in three and four dimensions, but not in two. Though the distinction is relatively clear when considering the relative values, the approach taken here will be to take such additional Gribov copies merely as an additional finite-volume effect, rather then attempting to treating them separately. Aside from this second peak, the peak at large $\Delta b$ is only very slightly dependent on volume and discretization, but shows a notable asymmetry towards smaller values.

Finally, the number of indistinct Gribov copies quickly diminishes with both larger volumes and finer discretizations, though in two dimensions this is still a large fraction. From now on, only Gribov copies distinct in both parameters will be considered, and be denoted by $n_g$ to distinguish them from the search space size $N_r$.

\subsection{Counting Gribov copies}

The second problem is that there are usually many Gribov copies expected. This is also observed here. Especially in larger volumes and higher dimensions, in most cases an increase in $N_r$ also increases the expectation value $N_g=\langle n_g\rangle$ of the number of found genuine copies $n_g$ per orbit. Therefore, to find the number of Gribov copies, the results should be extrapolated to $N_r\to\infty$.

\begin{figure}
\includegraphics[width=\linewidth]{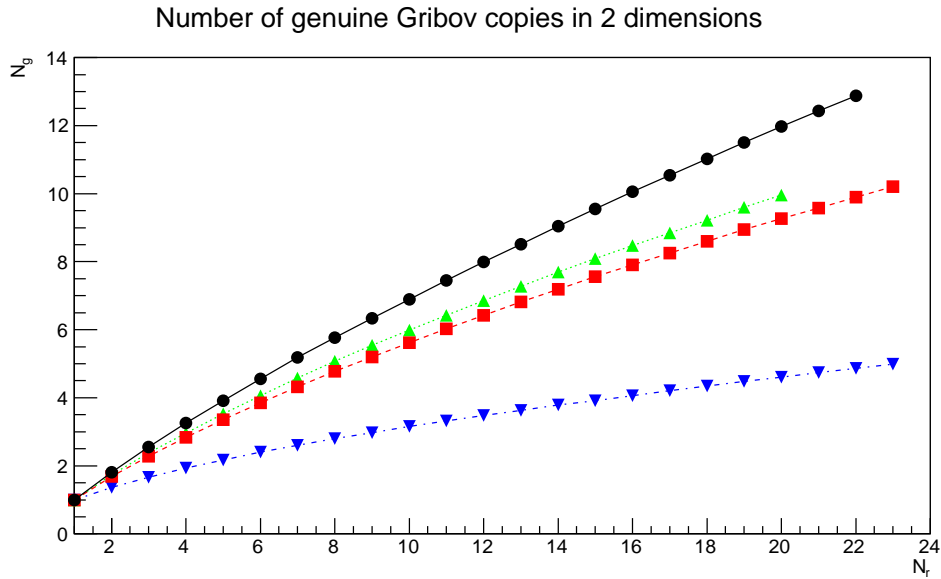}
\caption{\label{fig:2dngnr}Number of genuine Gribov copies $N_g$ as a function of the search space size $N_r$ in two dimensions. Symbols have the same meaning as in figure \ref{fig:2ddiff}. Error bars are smaller than the symbol sizes.}
\end{figure}

\begin{figure}
\includegraphics[width=\linewidth]{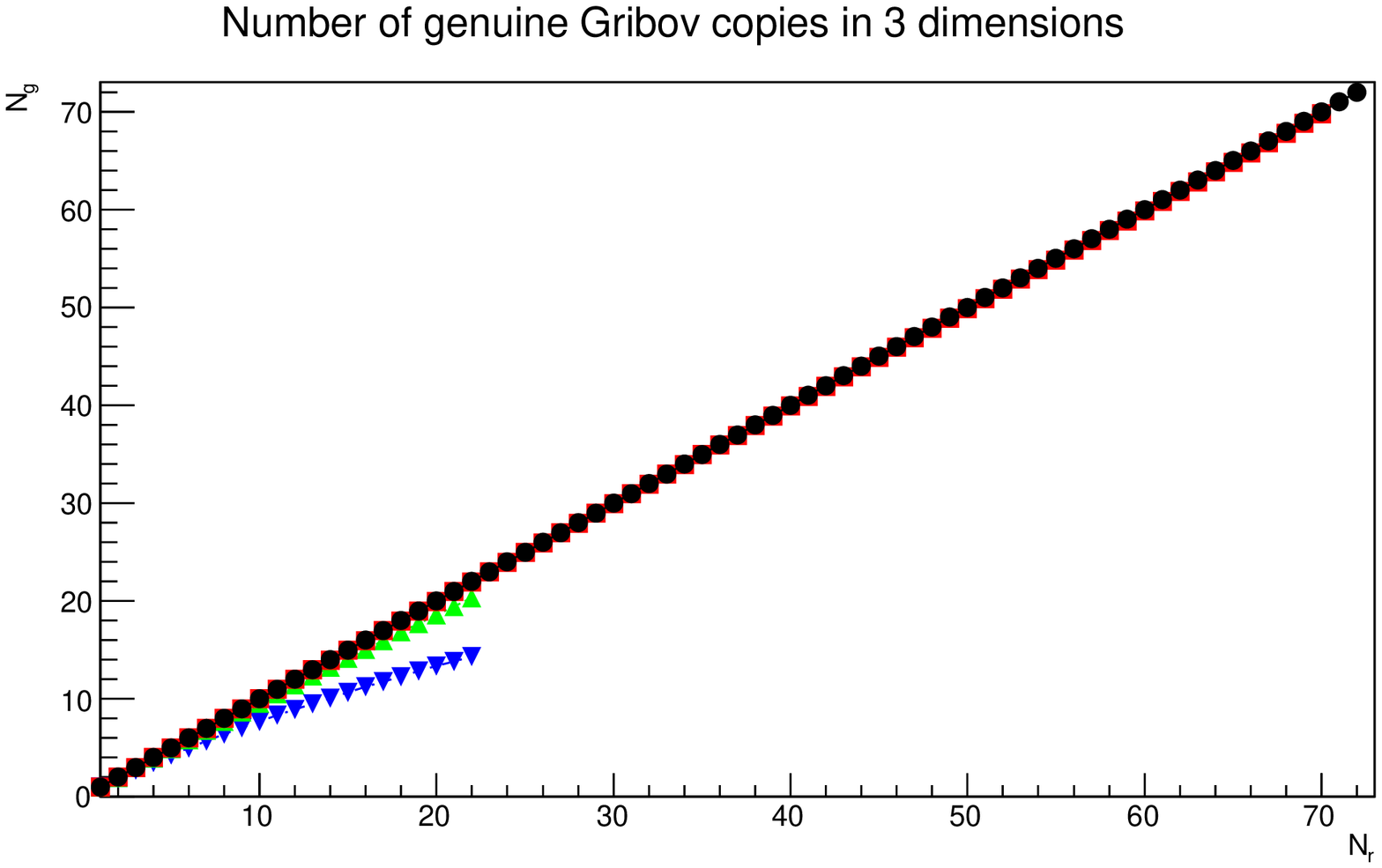}
\caption{\label{fig:3dngnr}Number of genuine Gribov copies $N_g$ as a function of the search space size $N_r$ in three dimensions. Symbols have the same meaning as in figure \ref{fig:3ddiff}. Error bars are smaller than the symbol sizes.}
\end{figure}

\begin{figure}
\includegraphics[width=\linewidth]{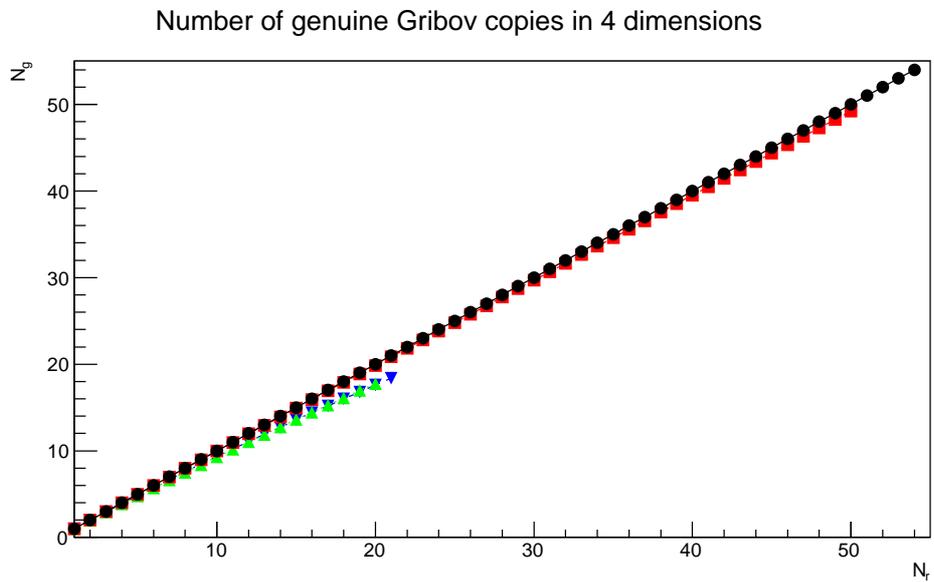}
\caption{\label{fig:4dngnr}Number of genuine Gribov copies $N_g$ as a function of the search space size $N_r$ in four dimensions. Symbols have the same meaning as in figure \ref{fig:4ddiff}. Error bars are smaller than the symbol sizes.}
\end{figure}

To investigate the dependence of $N_g$ on $N_r$, the results are shown in figures \ref{fig:2dngnr}-\ref{fig:4dngnr}. The results show that for sufficiently large volumes and higher dimensions more and more each increase of the search space size increases the number of genuine Gribov copies, yielding eventually an (almost) linear dependence. For smaller volumes and dimensions, the dependence is less than linear, and can be well fitted by power-laws, and thus shows no saturation with search space size. Furthermore, the exponents become closer and closer to one, the finer the discretization and the larger the volumes, approaching one. Thus, the extrapolation gives a linear dependency in the thermodynamic limit. Hence, no extrapolation is meaningful either at fixed lattice parameters nor to thermodynamic value, as they all yield an infinite number of Gribov copies. Thus, the only statement which can be made is that, even in small volumes, the number of genuine Gribov copies is very large. This is in line with results from attempts to explicitly count Gribov copies on tiny lattices \cite{Mehta:2011sp,Hughes:2012hg}.

\begin{figure}
\includegraphics[width=\linewidth]{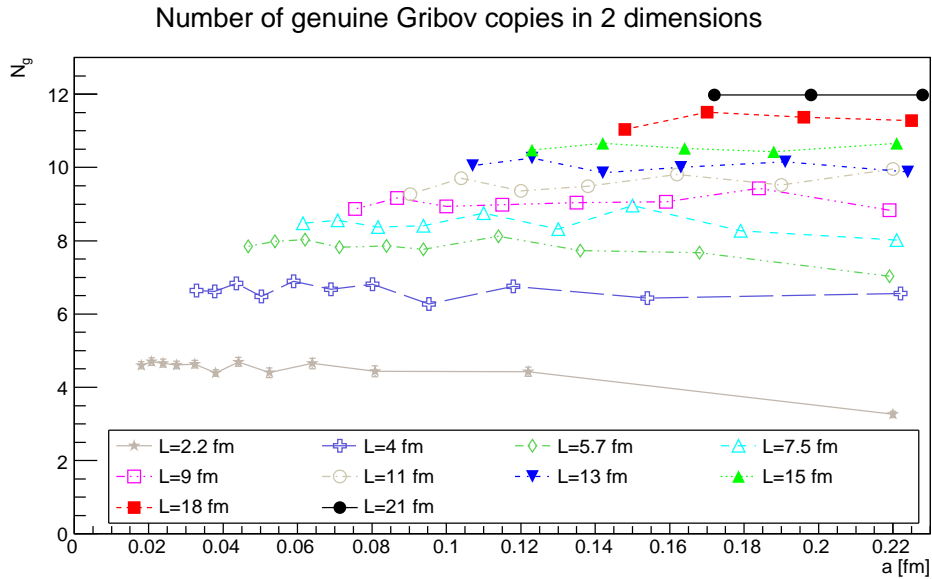}
\caption{\label{fig:2dngnra}Number of genuine Gribov copies $N_g$ at fixed search space size $N_r=20$ as a function of discretization and volume in two dimensions.}
\end{figure}

\begin{figure}
\includegraphics[width=\linewidth]{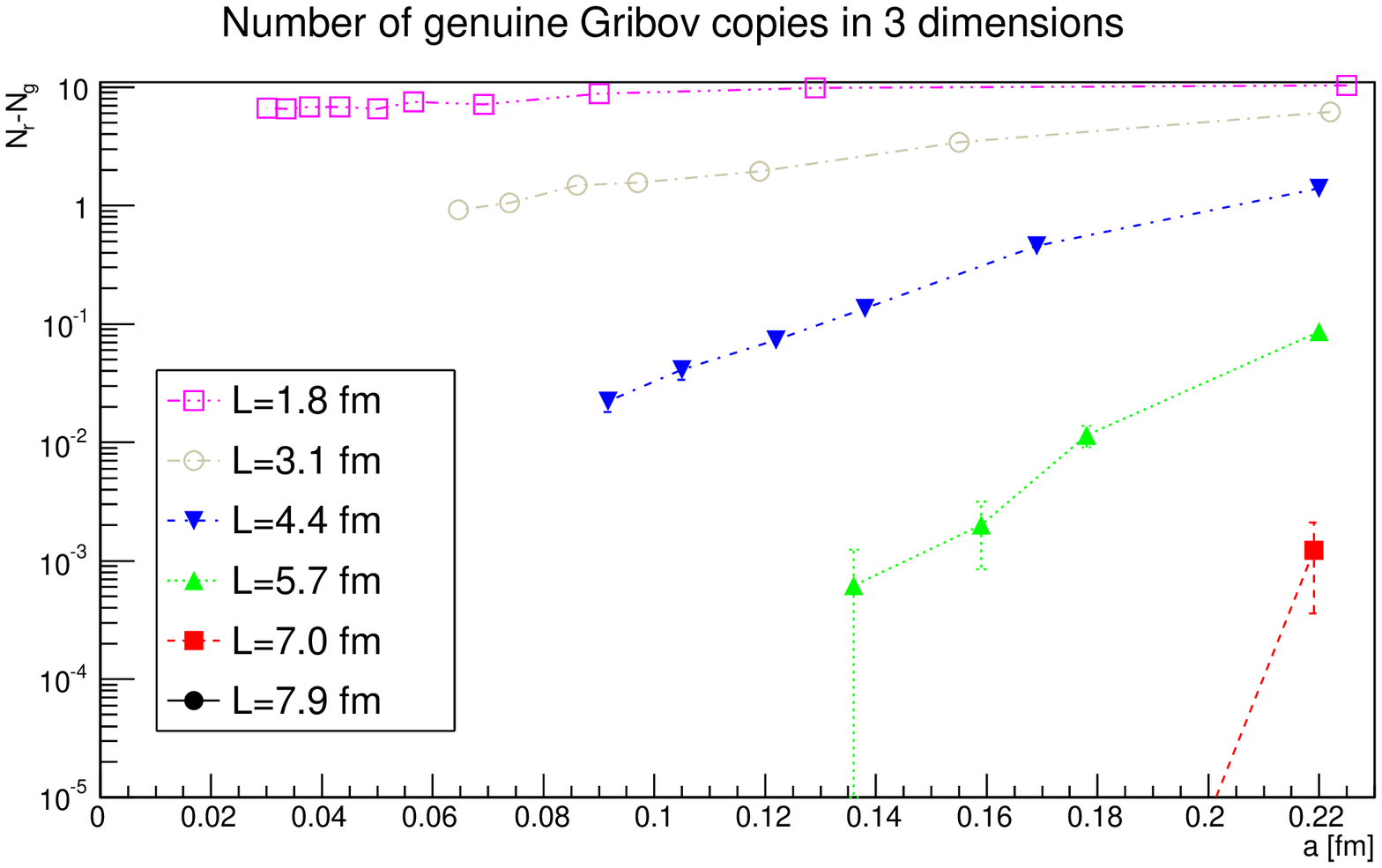}
\caption{\label{fig:3dngnra}Plot of $N_r-N_g$ at fixed search space size $N_r=20$ as a function of discretization and volume in three dimensions. Not displayed points are zero within numerical accuracy.}
\end{figure}

\begin{figure}
\includegraphics[width=\linewidth]{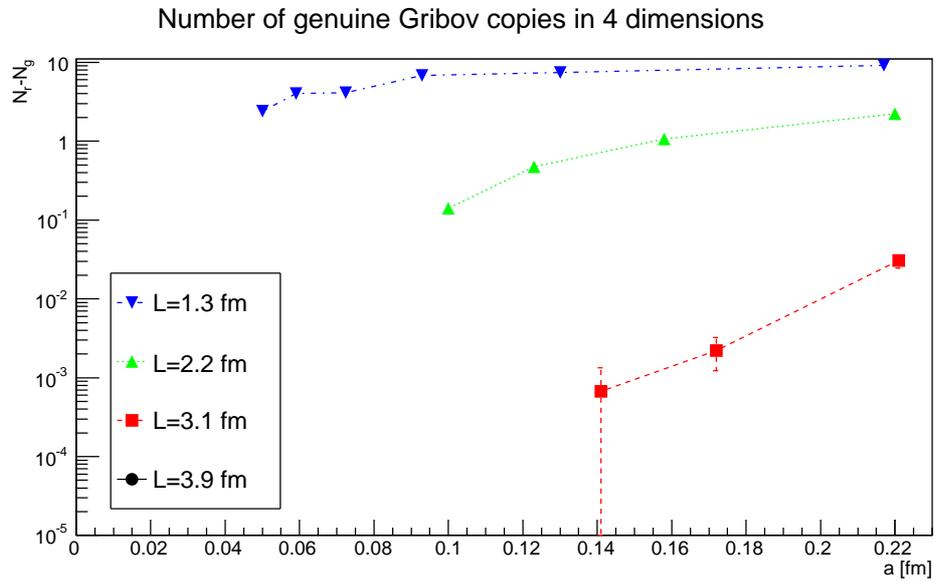}
\caption{\label{fig:4dngnra}Plot of $N_r-N_g$ at fixed search space size $N_r=20$ as a function of discretization and volume in four dimensions. Not displayed points are zero within numerical accuracy.}
\end{figure}

There is one remarkable observation, when investigating closer the dependence of the number of genuine Gribov copies found at a fixed search space size as a function of dimensionality, physical volume, and discretization, shown in figures \ref{fig:2dngnra}-\ref{fig:4dngnra}. It is visible that in two dimensions the number of genuine Gribov copies found is a slowly increasing function of volume, and only very weakly dependent, if at all, on the discretization. Especially at large volumes, the number is essentially fixed as a function of lattice discretization. This is not true in three and four dimensions, where the number of genuine Gribov copies quickly increases both with physical volume and lattice discretization. Especially at large volumes every new Gribov copy is distinct, even on very coarse lattices. This is even more pronounced in four dimensions than in three dimensions. This indicates that the first Gribov region in two dimensions may be different from the ones in three and four dimensions. This would add to the list of features in which two dimensions appears to be different from higher dimensions \cite{Maas:2011se}. However, since Gribov copies are a pure gauge effect, it seems unlikely to be related to the absence of dynamics in two dimensions. Whatever the reason, it has likely to be something which is entirely due to the gauge structure.

\begin{figure}
\includegraphics[width=\linewidth]{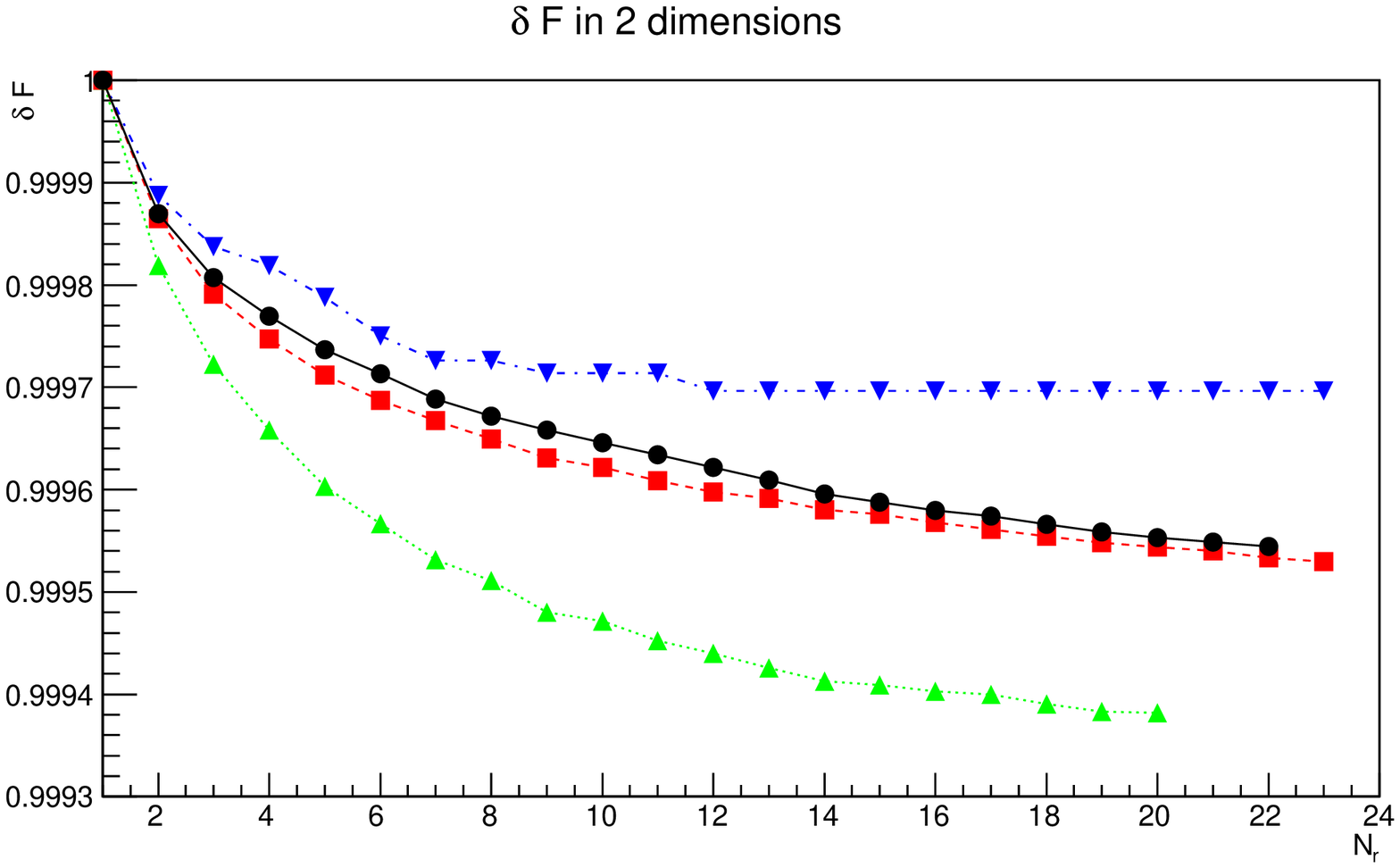}\\
\includegraphics[width=\linewidth]{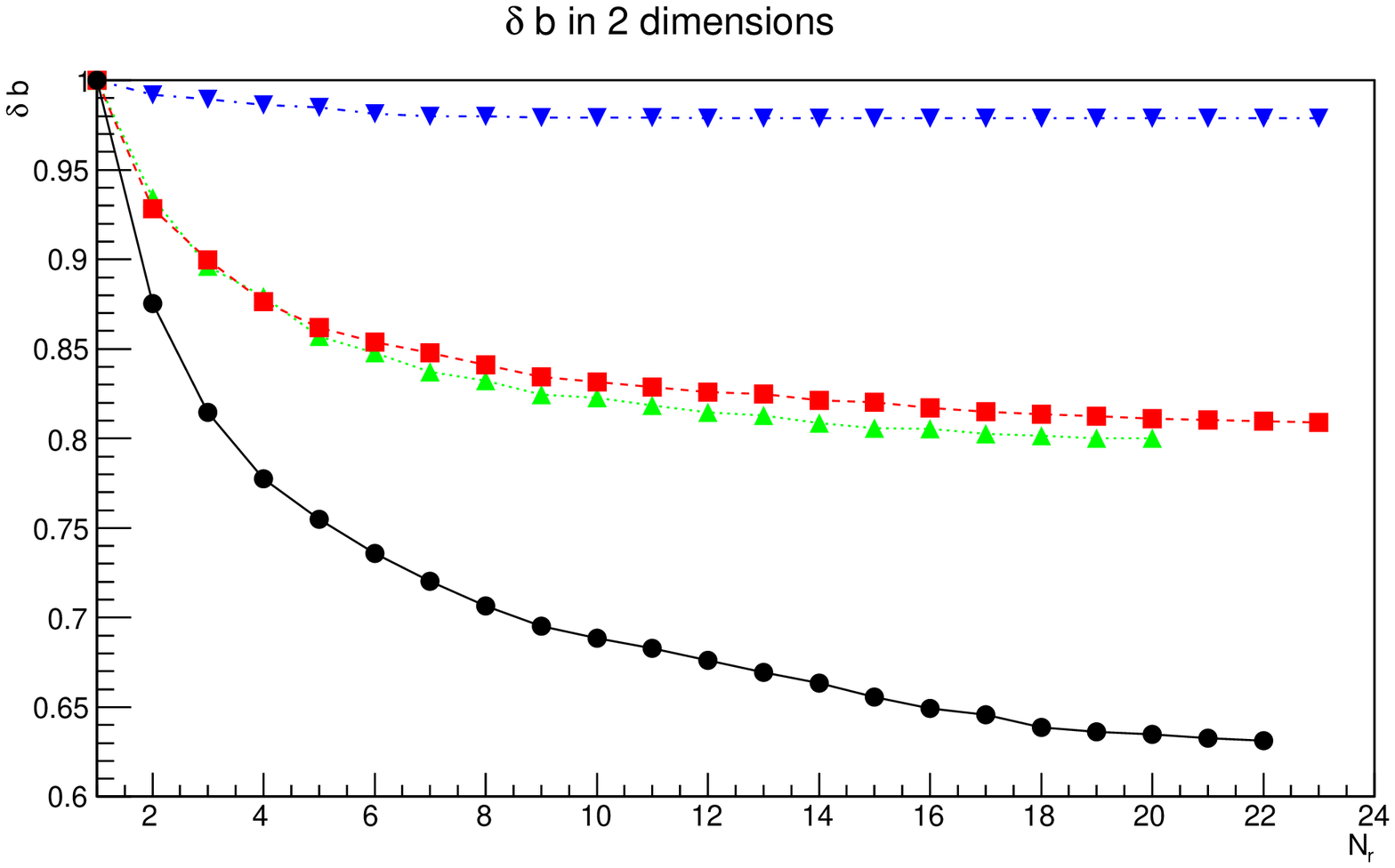}
\caption{\label{fig:2ddelta}The discrimination parameter corridor width $\delta F$ (top panel) and $\delta b$ (bottom panel) in two dimensions. Symbols have the same meaning as in figure \ref{fig:2ddiff}. Error bars are smaller than the symbol sizes.}
\end{figure}

\begin{figure}
\includegraphics[width=\linewidth]{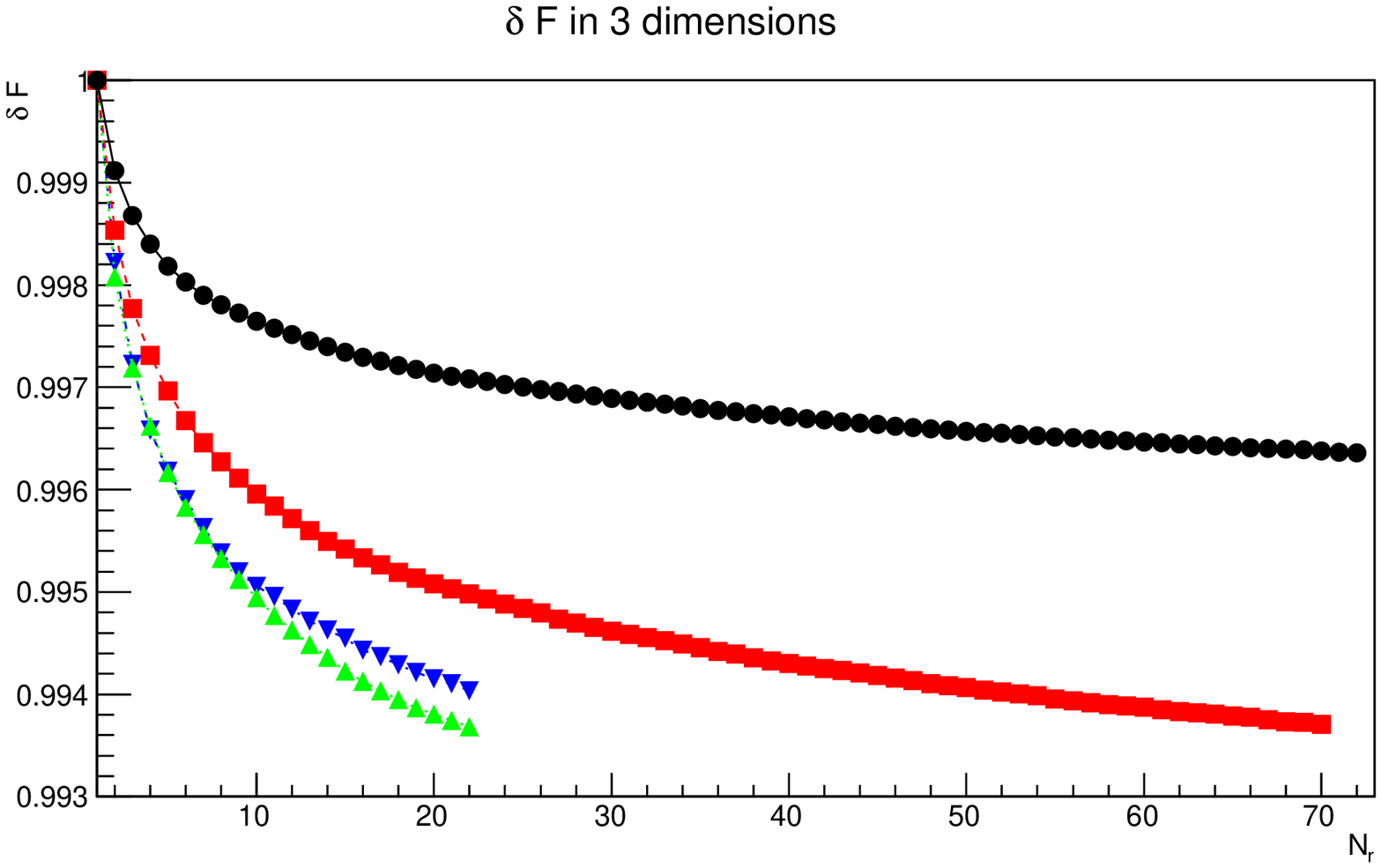}\\
\includegraphics[width=\linewidth]{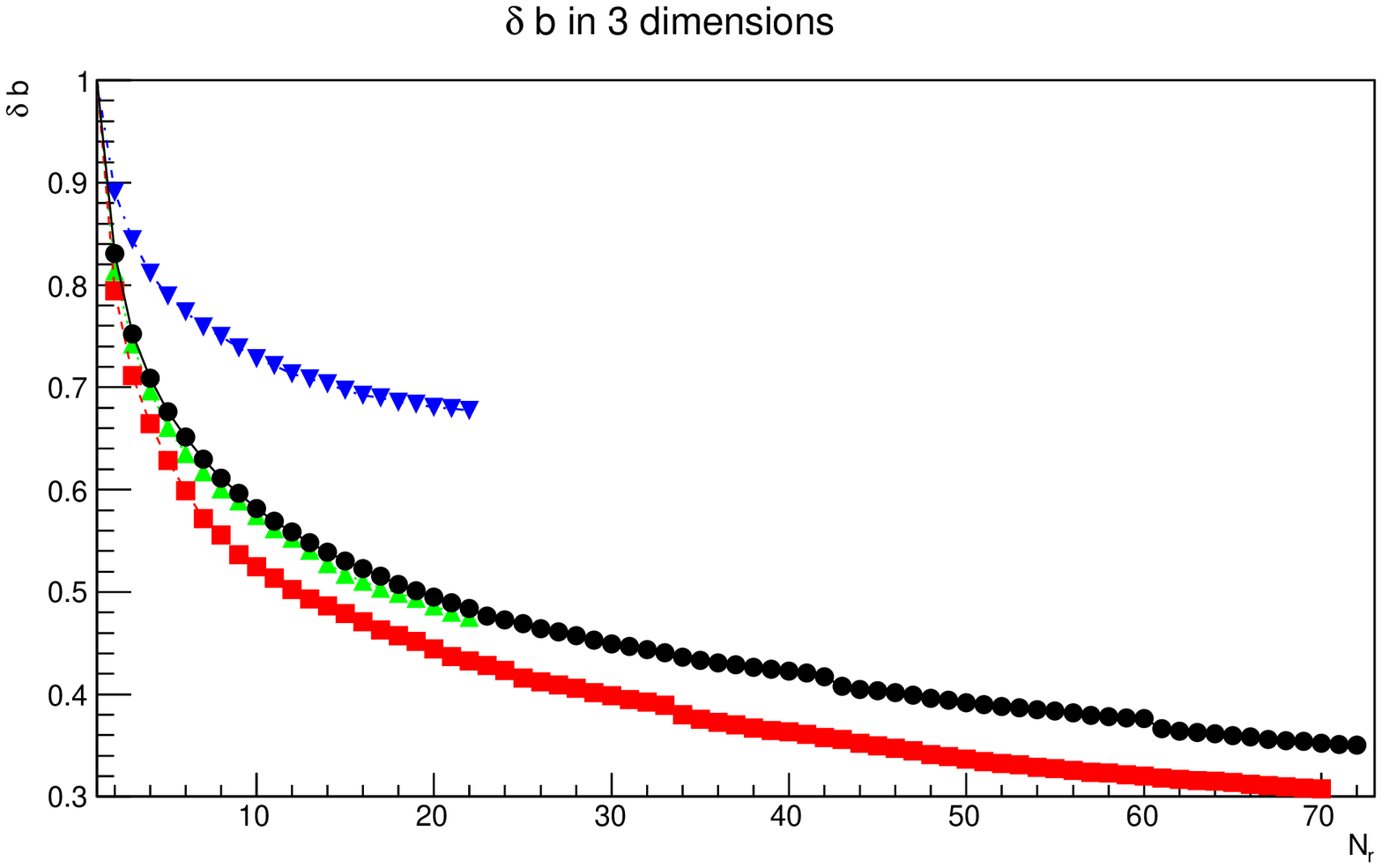}
\caption{\label{fig:3ddelta}The discrimination parameter corridor width $\delta F$ (top panel) and $\delta b$ (bottom panel) in three dimensions. Symbols have the same meaning as in figure \ref{fig:3ddiff}. Error bars are smaller than the symbol sizes.}
\end{figure}

\begin{figure}
\includegraphics[width=\linewidth]{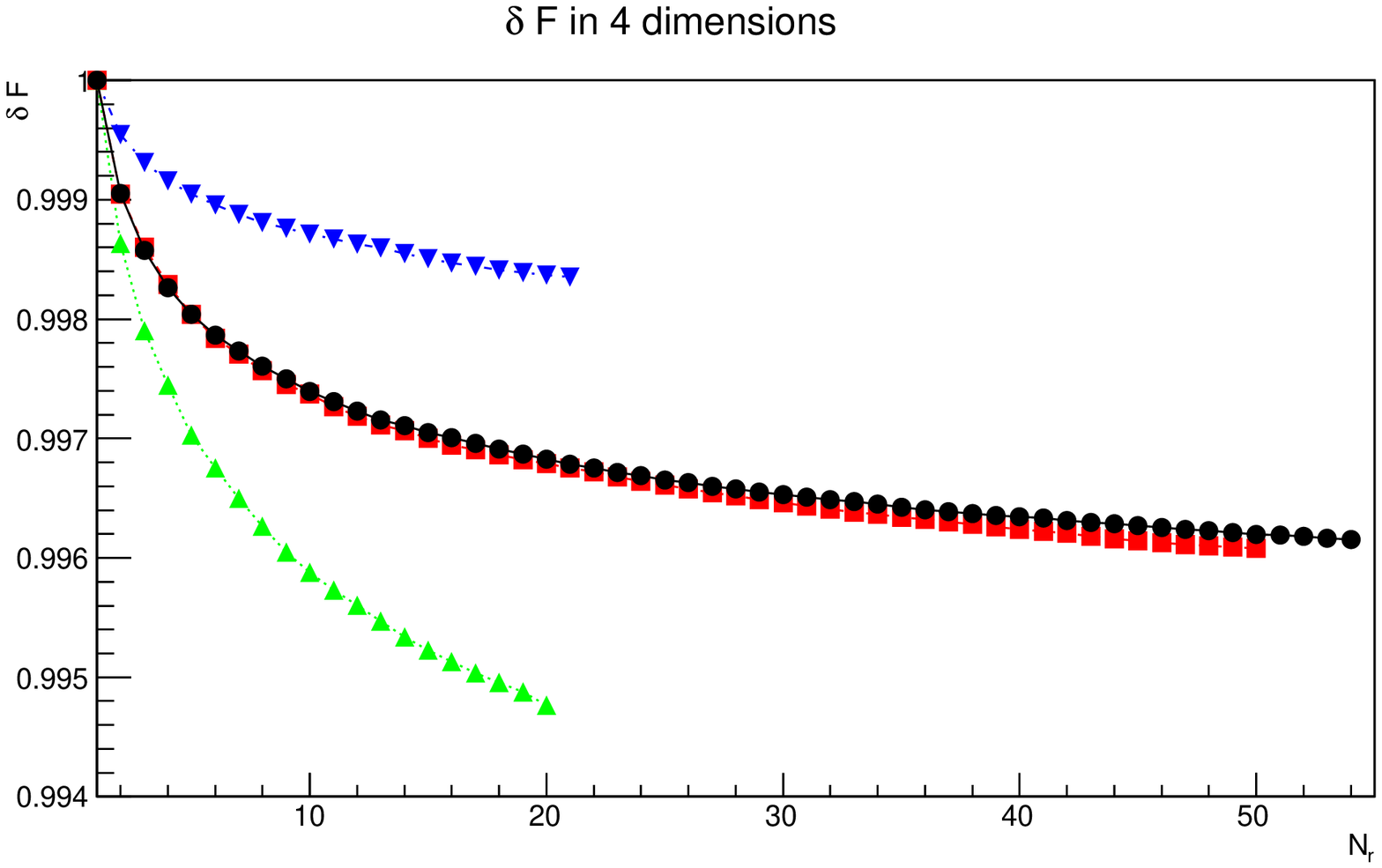}\\
\includegraphics[width=\linewidth]{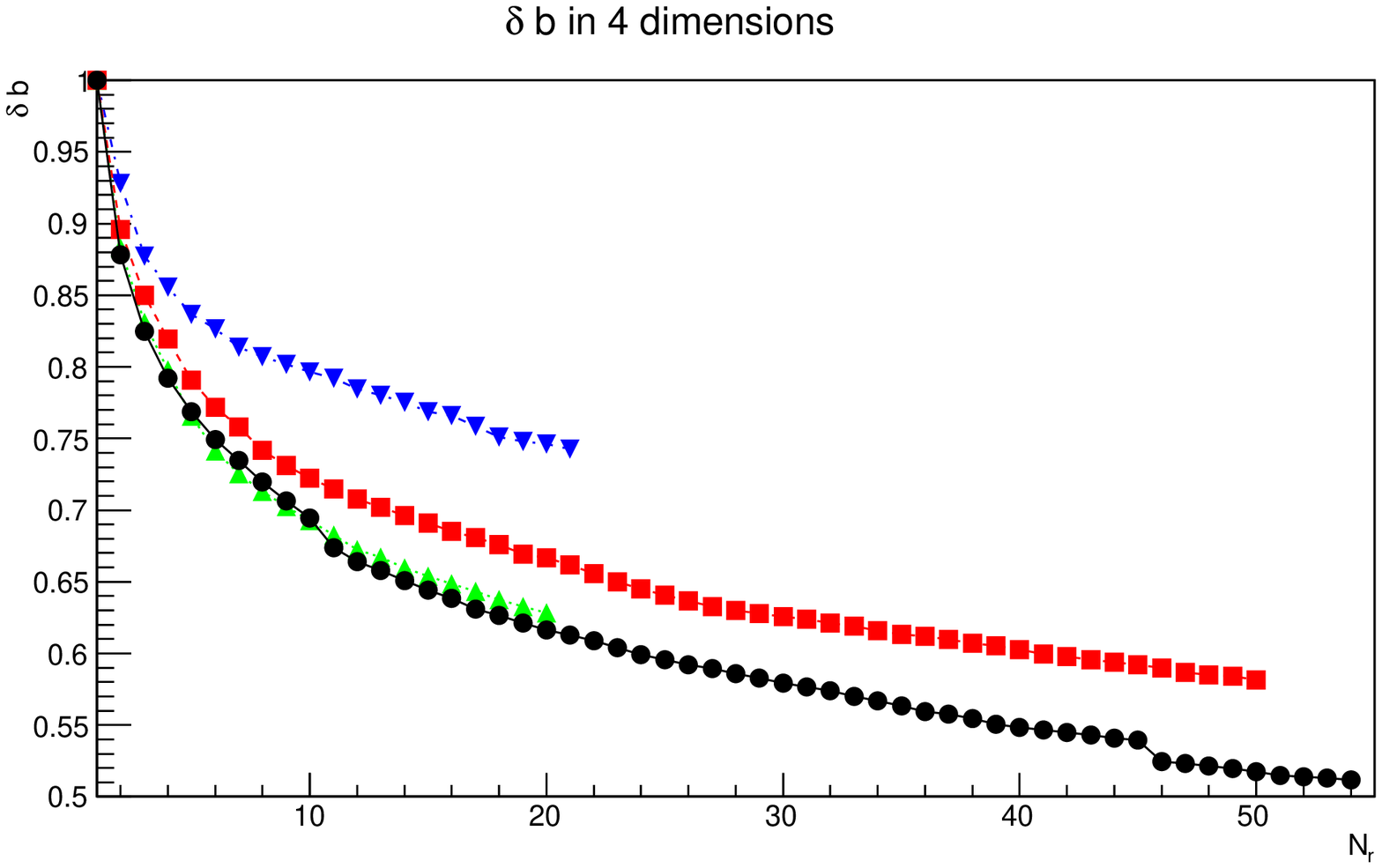}
\caption{\label{fig:4ddelta}The discrimination parameter corridor width $\delta F$ (top panel) and $\delta b$ (bottom panel) in four dimensions. Symbols have the same meaning as in figure \ref{fig:4ddiff}. Error bars are smaller than the symbol sizes.}
\end{figure}

The question is, whether the impossibility to, even by extrapolation, count the number of Gribov copies limits the possibility to assess the impact of Gribov copies. Especially, if gauge-fixed quantities can nonetheless be extrapolated, or whether it is just possible to give lower limits to their variation under increase of the search space size. To investigate this problem, as an example in figures \ref{fig:2ddelta}-\ref{fig:4ddelta} the quantities
\be
\delta D=\frac{\langle\min D\rangle}{\langle\max D\rangle}\nn,
\ee
\no that is the width of the discrimination parameter corridor, are shown. Again, the ratio ensures that all trivial factors drop out. The results show initially a very strong dependence on the search space size, but then start to flatten out. Above $N_r\approx 10$, the behavior can be fitted rather well with the function
\be
f(N_r)=1-\frac{N_r^a}{b+c N_r^a}\overset{N_r\to\infty}{=}1-\frac{1}{c}\label{fit},
\ee
\no which has a finite limit. Therefore an extraction of meaningful quantities is possible, even if the search space is not large enough to get close to the asymptotic value. In fact, approaching the asymptotic value of 0.9950 up to 0.9955(0.9951) for $L=3.9$ fm at $a=0.22$ fm in four dimensions for $\delta F$ would require, according to the fit, $N_r\approx 220(2500)$, which is rather large. The same functional form and the same statements also apply to $\Delta b$.

Interestingly, occasionally small jumps are seen. These originate if a single Gribov copy has a strong impact on the final result, indicating that the statistics at this search space size is, in principle, not large enough. These exceptional Gribov copies \cite{Cucchieri:2006tf,Sternbeck:2005tk,Maas:2008ri} originate from the fact that the distributions in figure \ref{fig:2ddiff}-\ref{fig:4ddiff} are non-Gaussian and have long tails. This effect diminishes with an increase in the number of genuine Gribov copies, and thus on larger volumes and finer discretizations, at least in more than two dimensions. Nonetheless, this shows that it is, in principle with sufficient statistic and a finites search-space size, possible to find meaningful results, and not only limits.

\begin{figure}
\includegraphics[width=0.5\linewidth]{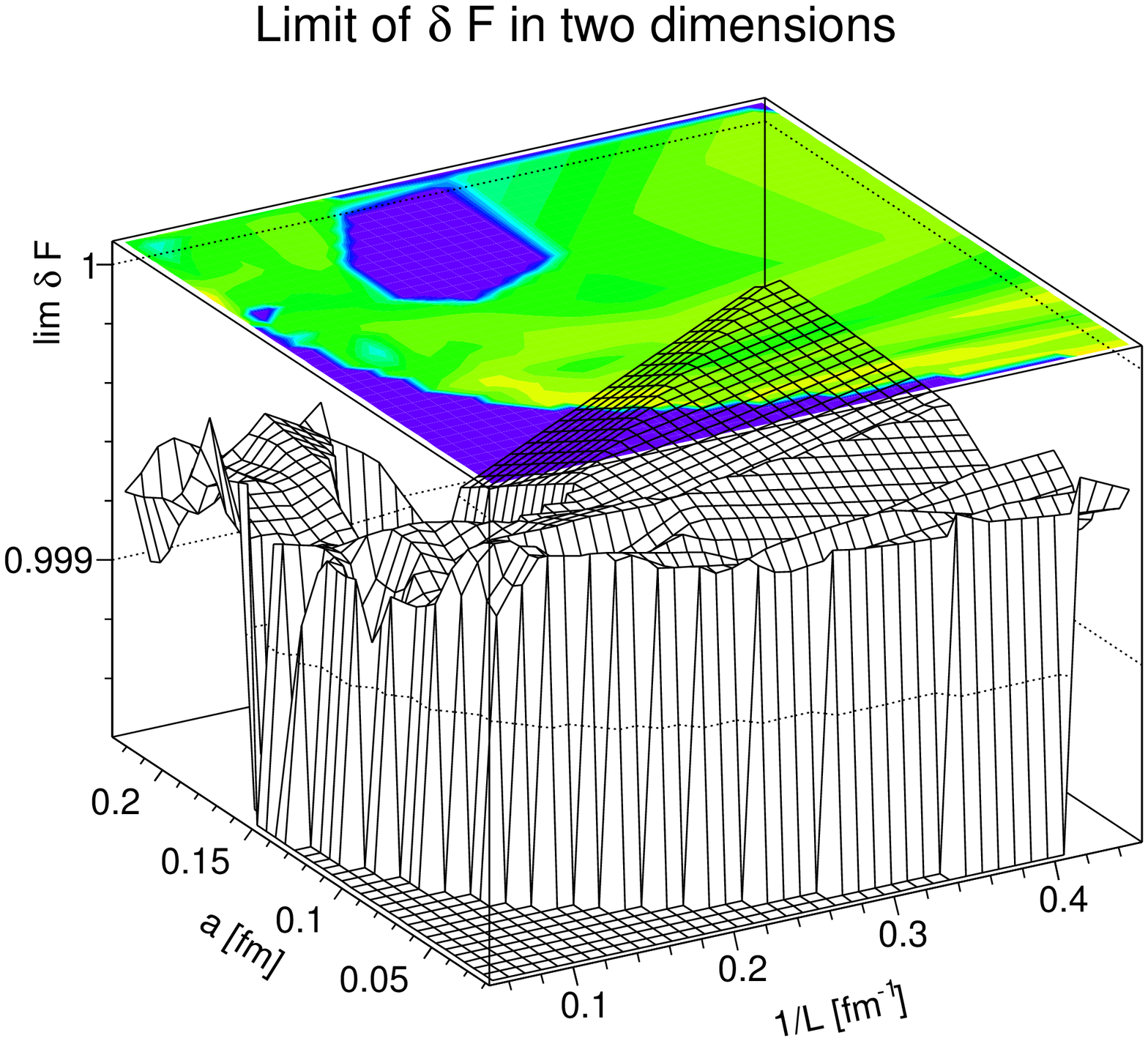}\includegraphics[width=0.5\linewidth]{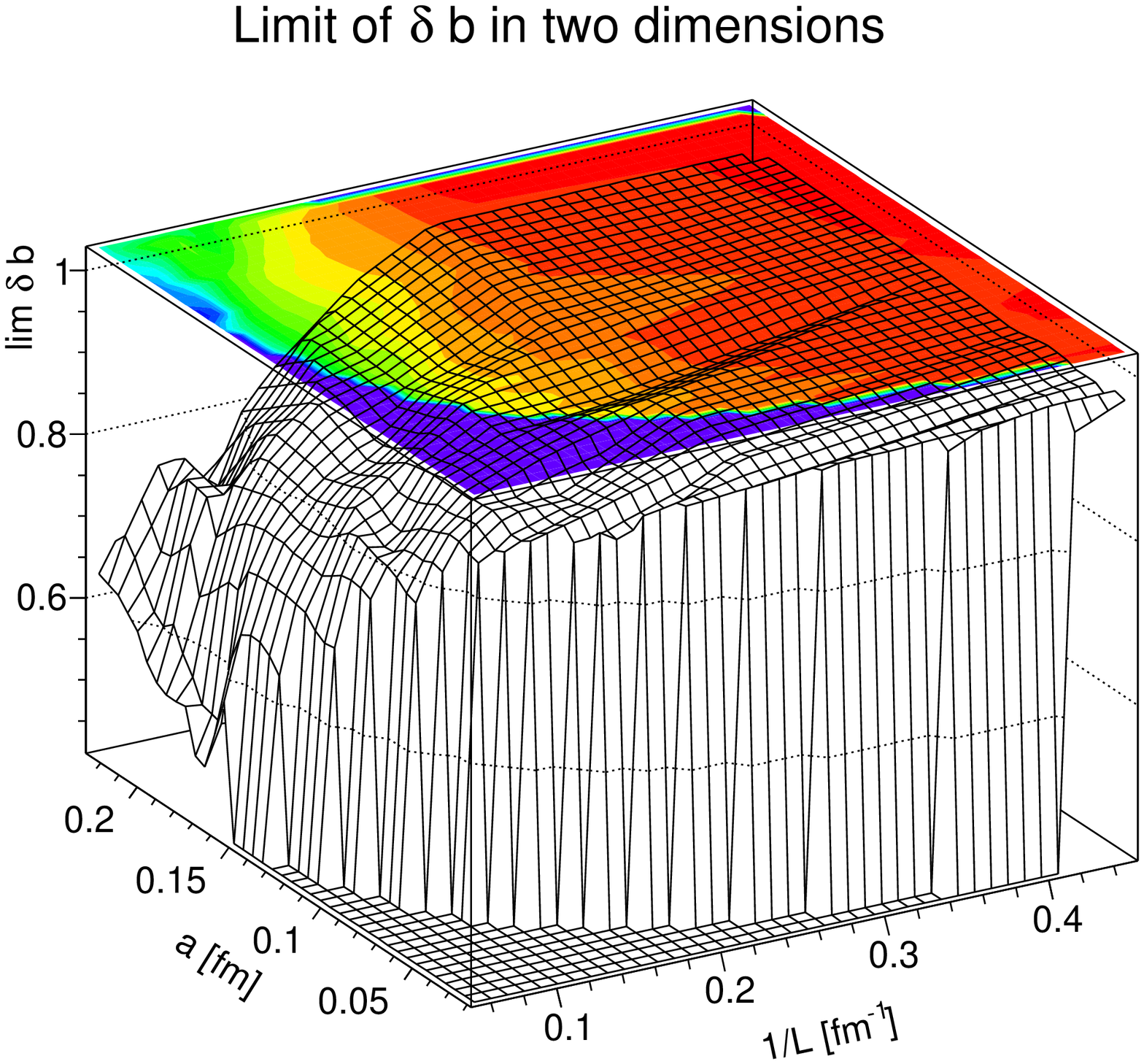}\\
\includegraphics[width=0.5\linewidth]{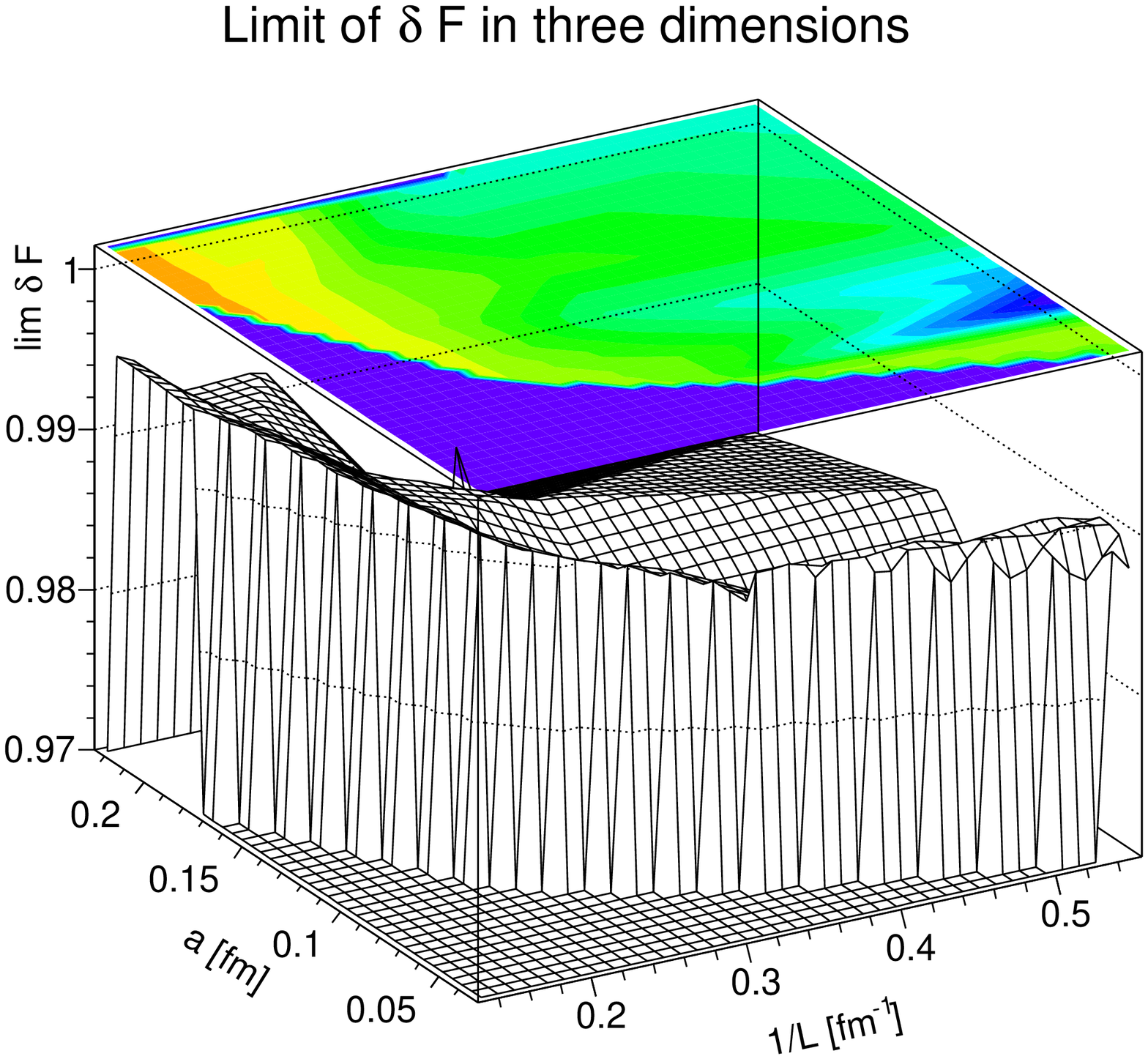}\includegraphics[width=0.5\linewidth]{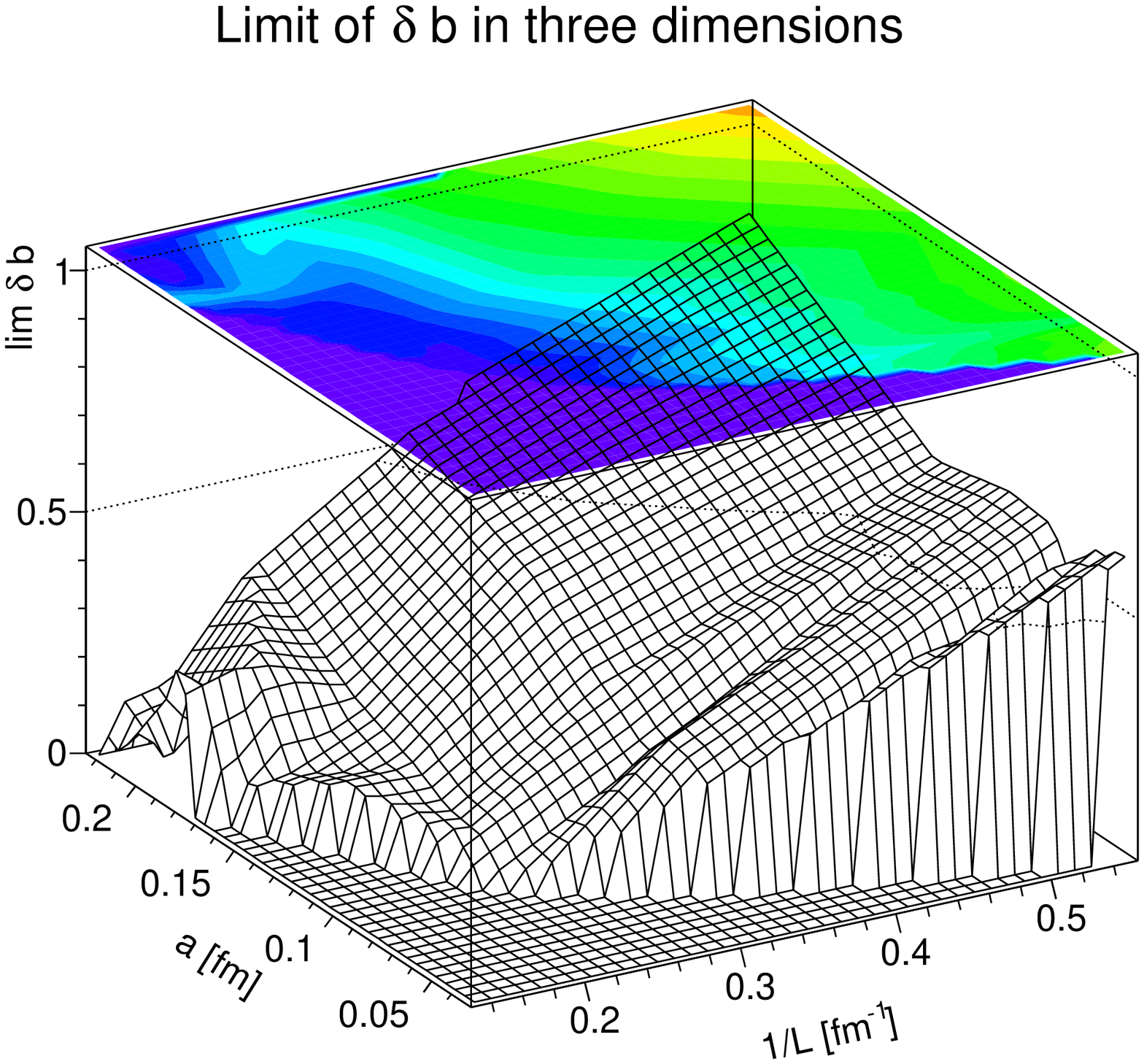}\\
\includegraphics[width=0.5\linewidth]{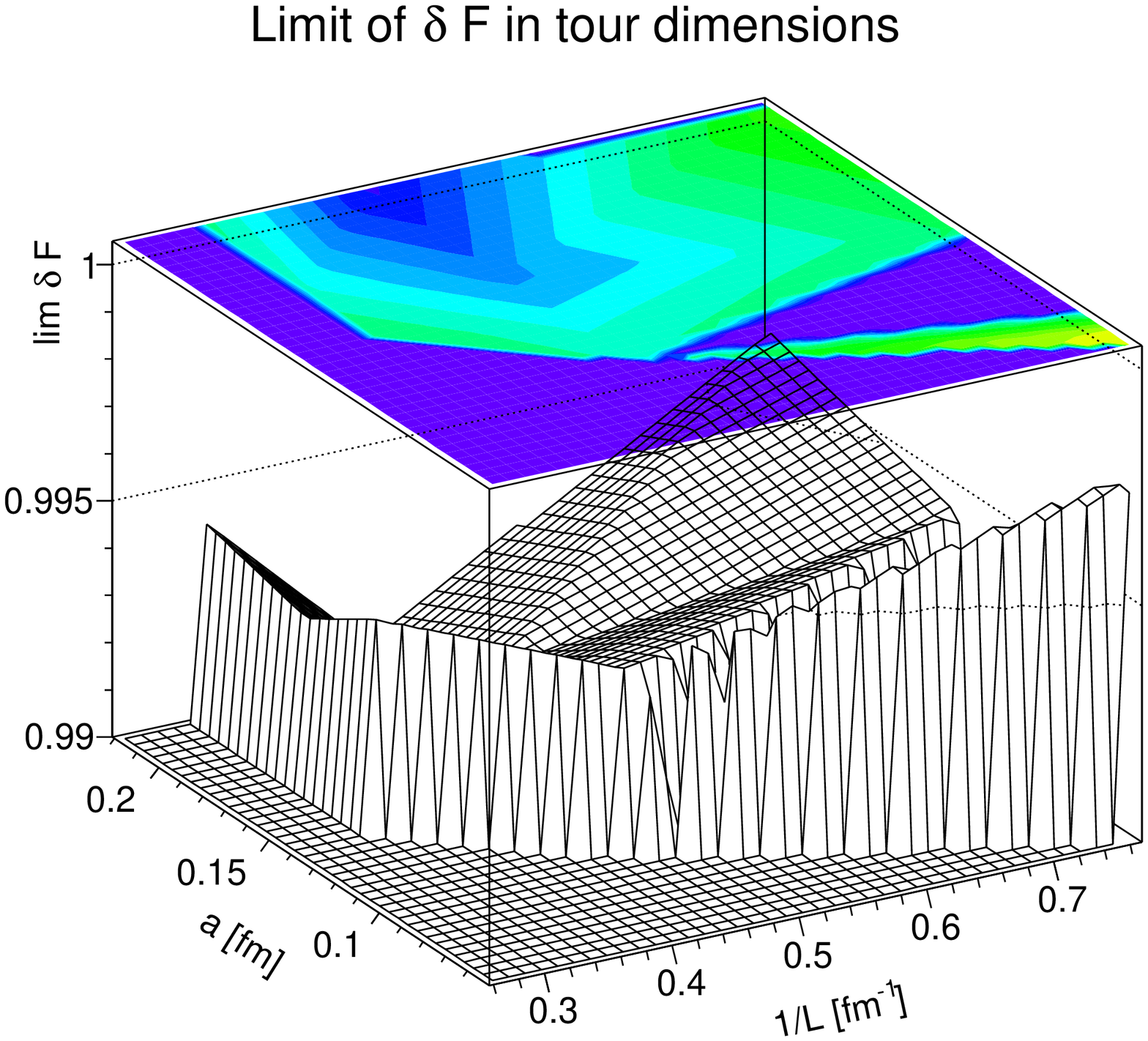}\includegraphics[width=0.5\linewidth]{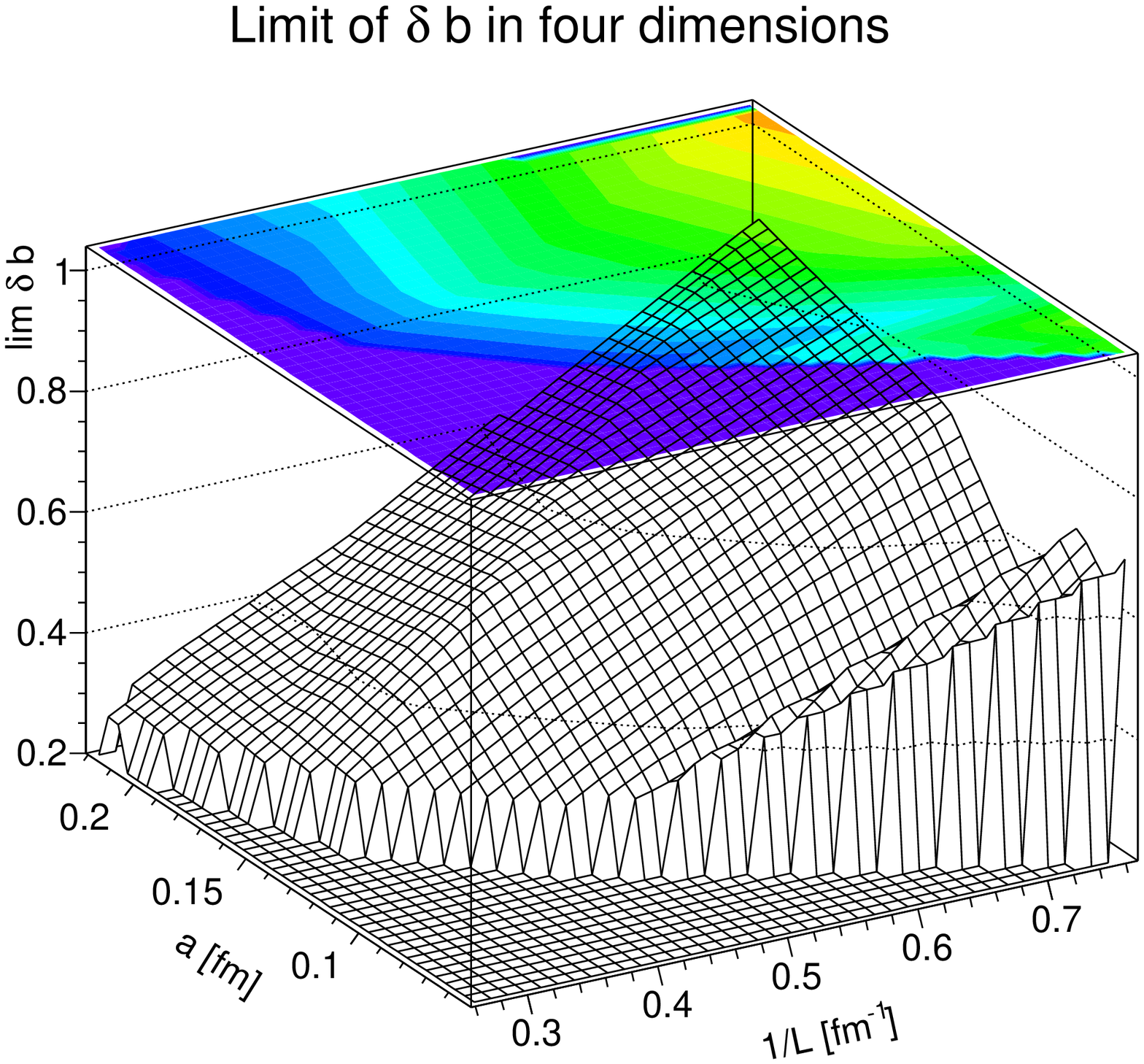}\\
\caption{\label{fig:limits}. The limits of $\delta F$ (left panels) and $\delta b$ (right panels) as obtained with \pref{fit} as a function of physical extent and lattice spacing in two dimensions (top panels), three dimensions (middle panels), and four dimensions (bottom panels).}
\end{figure}

To continue the example, the limits obtained from \pref{fit} as a function of physical volume and lattice spacing are shown in figure \ref{fig:limits}. The strongest systematic effects in the fits occur due to the exceptional configurations, giving rise to most strong deviations from the trend. This surfaces in the plots of figure \ref{fig:limits} as occasional dips. This problem is strongest in two dimensions, where the number of genuine Gribov copies is comparatively small, and a single exceptional configuration can have a large impact.

While the limit of $\delta F$ does show some indication of tending to 1 at large volumes and finite discretizations, and thus indicating indeed that the minima become more degenerate as in Abelian gauge theory \cite{deForcrand:1994mz} and as expected due to the general arguments \cite{Zwanziger:1993dh,Cucchieri:2013xka}, the situation for $\delta b$ is quite different. In this case, $\delta b$ tends to smaller values for larger and, in three and four dimensions, finer lattices. However, the results do not permit a stable extrapolation, and therefore it is impossible to decide whether it may tend to zero or not. However, in three and four dimensions a value of zero, indicating a divergence in $\langle \max b\rangle$, is not favored in the current understanding of the first Gribov region as well as from direct calculations of the ghost dressing function \cite{Fischer:2008uz,Maas:2011se,Cucchieri:2008fc,Cucchieri:2013nja}. But in two dimensions this appears to be the case \cite{Maas:2007uv,Maas:2014xma,Cucchieri:2008fc,Dudal:2008xd,Dudal:2012td,Huber:2012td}.

\subsection{Orbit dependence}

A major question for constructing complete Landau gauges is whether the properties of the residual gauge orbit are gauge-orbit-independent \cite{Maas:2009se,Maas:2011se,Maas:2013vd,Serreau:2012cg}. Especially, consider some expectation value
\be
\langle{\cal O}\rangle=\frac{\int D A D g{\cal O} e^{i(S+S_{\text{gf}})}}{\int D A D g e^{i(S+S_\text{gf})}}\overset{{\cal O}\text{ gauge-invariant}}{=}\frac{\int D A {\cal O} e^{iS}\int Dg e^{iS_\text{gf}}}{\int D A e^{iS}\int Dg e^{iS_{\text{gf}}}}\nn,
\ee
\no where $A$ denotes the integral over gauge orbits and $g$ the integral over all admitted Gribov copies of the specified residual gauge orbits. For any permissible gauge, the separated integral over the gauge-fixing condition has to cancel, at least formally, for a gauge-invariant observable. However, if the integral becomes orbit-dependent, then in the expectation value every factor from the gauge-fixing part is weighted by the observable\footnote{If ${\cal O}=1$, it would still work, the problem arises if the observable becomes orbit-dependent.}, and therefore the result may not cancel, yielding an invalid gauge condition. Thus an admissible gauge condition must have on every orbit (up to a measure-zero contribution) the same weight, such that the gauge-fixing factor can be pulled out. This can be either achieved by having a gauge-condition which has intrinsically the same weight on every orbit or by having an orbit-dependent normalization factor to achieve this\footnote{Note that additional complications may arise if the ability of the algorithm used in a lattice calculation to find Gribov copies would depend on the properties of the gauge orbits. This complication will be ignored here.}. For gauge-dependent quantities, this problem is present as well, but there it would still be possible to define this as part of an (odd) gauge-fixing prescription. But for gauge-invariant observables, there is no choice.

\begin{figure}
\includegraphics[width=\linewidth]{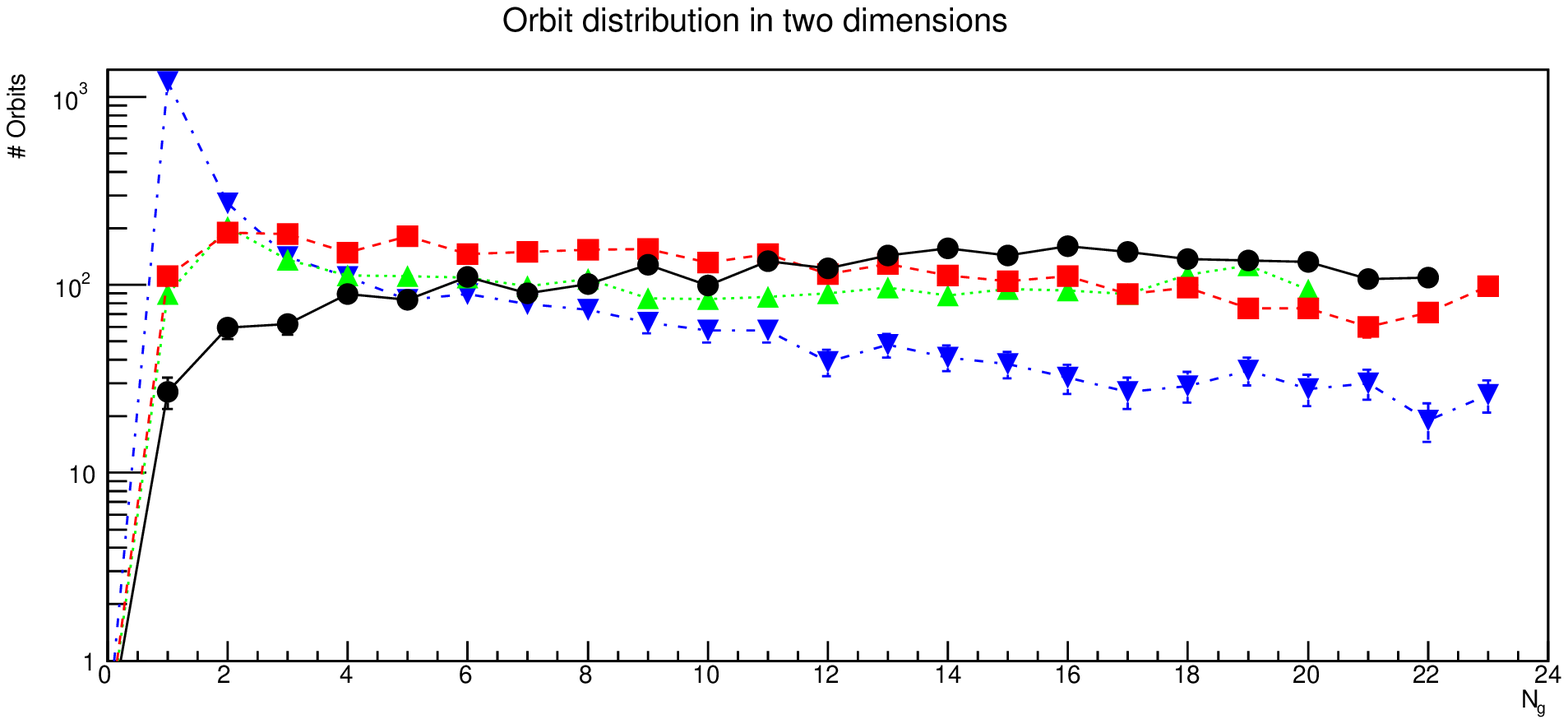}\\
\includegraphics[width=\linewidth]{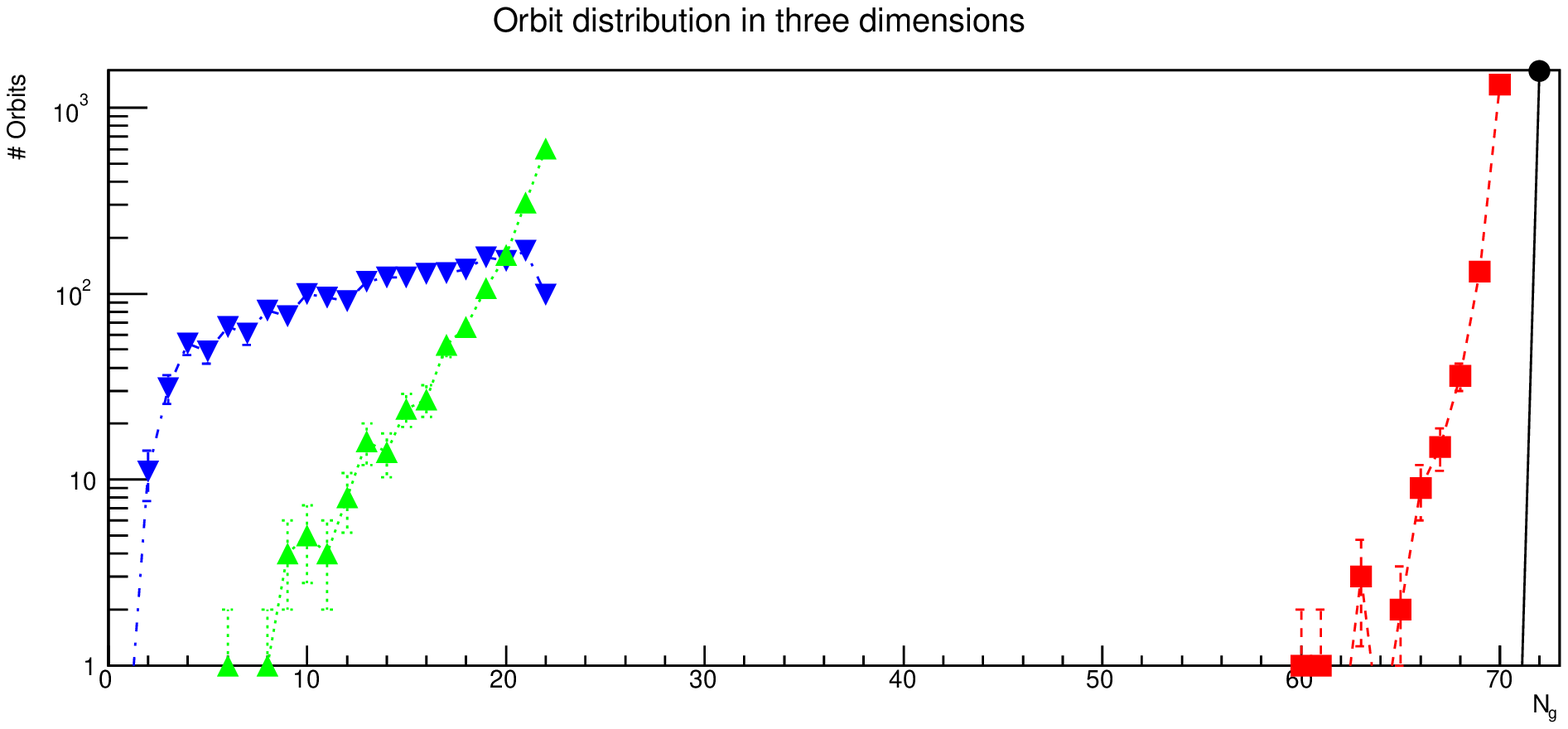}\\
\includegraphics[width=\linewidth]{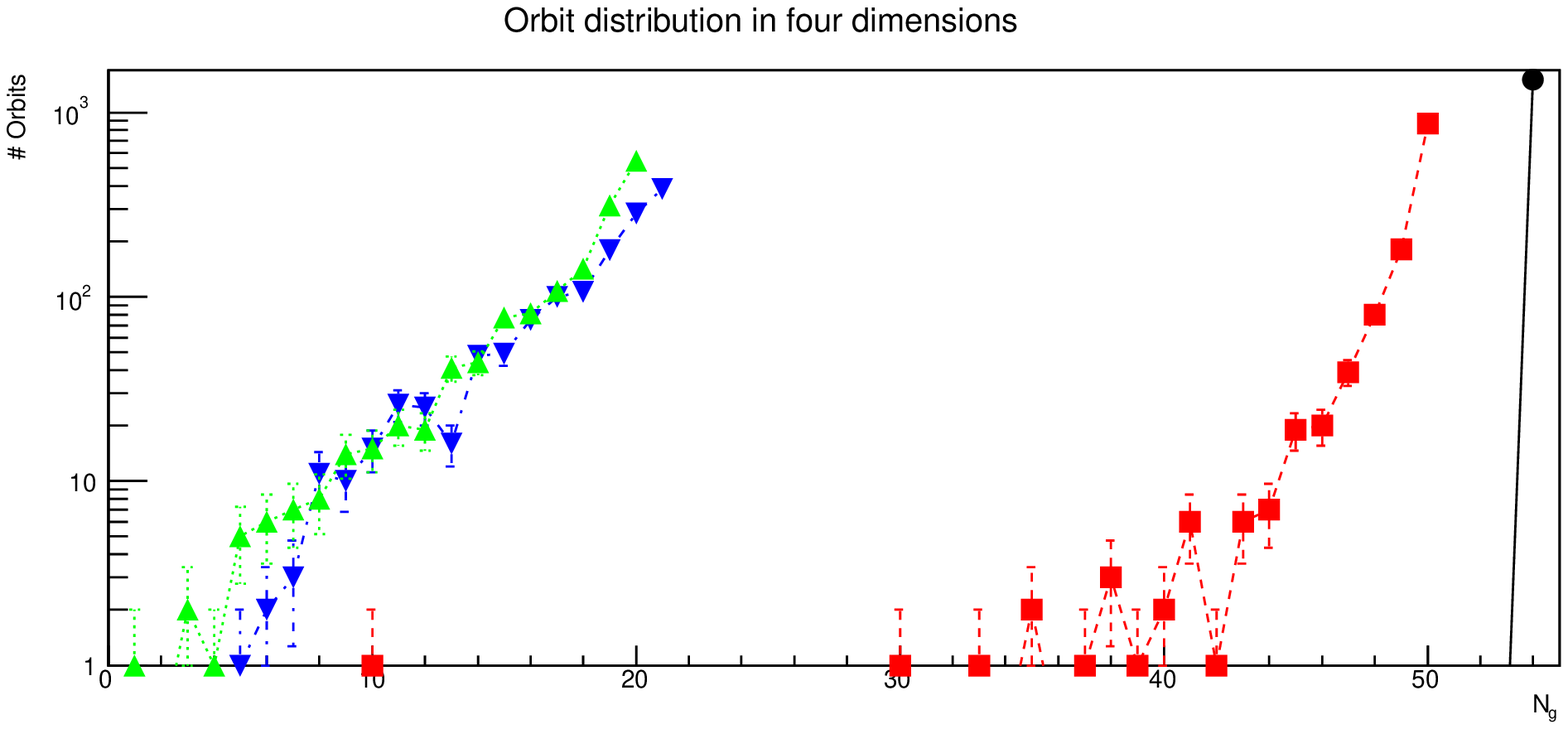}\\
\caption{\label{fig:orbit}. The number of orbits with the same number of genuine Gribov copies in two dimensions (top panel), three dimensions (middle panel), and four dimensions (bottom panel). Symbols are the same as in figures \ref{fig:2ddiff}-\ref{fig:4ddiff}.}
\end{figure}

Any gauge condition built upon selecting Gribov copies will be intrinsically orbit-independent if it has the same integrated weight on every orbit and every orbit has the same number of Gribov copies. To estimate the situation, the distribution of genuine Gribov copies for different orbits is shown in figure \ref{fig:orbit}. In two dimensions, the resulting structure is, except for the smallest lattices, extremely broad, with a maximum wandering towards larger number of genuine copies. The result is thus far from conclusive. In higher dimensions, the distribution becomes rather quickly narrower. However, the maximum quickly tends to the size of the search space, as seen previously. It can therefore not be excluded that the distribution would remain broad if the search space were larger. Thus, the results are not really conclusive also in higher dimensions. The only thing that can be stated is that with the given search space size there is at least no explicit disagreement with the hypothesis that in the thermodynamic limit the number of Gribov copies on different orbits only differ by an irrelevant number. But it is clear that on a finite lattice, and especially in two dimensions, the orbit dependence must be taken duly into account if a gauge should average over the residual gauge orbit\footnote{Note that minimal Landau gauge, which averages over the whole gauge orbit with a flat space by picking a random Gribov copy is not affected by this problem, as the averaging over the number of copies is automatically correct when only selecting a single representative. That is the same mechanism why finite-statistics lattice simulation do not need to worry about the size of the gauge orbit for gauge-invariant observables, where it is tacitly assumed that all gauge orbits have the same size, i.\ e.\ the same number of Gribov copies up to a measure zero difference.}.

\section{Structure of the first Gribov region}\label{s:struct}

After having now statements about the Gribov copies, this second part focuses on the properties of the Gribov region. Three separate features will be addressed, which have been investigated in the past. These are the boundary of the Gribov region, the so-called Gribov horizon, the interior of the region, and finally the fundamental modular region.

\subsection{The boundary of the first Gribov region}

One subject which has been of particular interest \cite{Zwanziger:1993dh,Cucchieri:2013nja,Greensite:2010hn,Greensite:2004ur,Cucchieri:2013xka} has been the boundary of the first Gribov region, i.\ e.\ the Gribov copies for which the Faddeev-Popov operator has a zero eigenvalue \cite{Gribov:1977wm}. 

This boundary is expected to have a non-regular shape \cite{Zwanziger:1993dh,Cucchieri:2013nja,Greensite:2010hn,Greensite:2004ur}. The actual Gribov horizon has as elements field configurations, and therefore it is defined in an infinite-dimensional space. Here, it will be again projected down to a two-dimensional one, parameterized by the two parameters\footnote{Note that in very few cases, usually at most one Gribov copy in the entire statistics for a given lattice setup, the inversion used to determine $b$ \cite{Cucchieri:2006tf} did not converge and yielded a negative value of $b$. These cases have been dropped here.} $b$ and $F$, now evaluated on single configurations. Since finite-dimensional spaces have different properties than infinite-dimensional ones, and especially two-dimensional ones, this may yield substantially different structures.

In a finite volume, the Gribov region is actually smaller than the one at infinite-volume, as in a finite volume the Faddeev-Popov operator has no genuine zero modes. To identify the actual boundary thus requires to assume that the lowest eigenvalue at finite volume is also signifying the Gribov copy which will become the boundary element in an infinite volume, and not, e.\ g., the one at the next-to-smallest. This corresponds to the assumption of the absence of level crossing. Especially on large volumes, this appears to be a reasonable assumption.

The next problem is that $b$ is not in one-to-one correlation with this smallest eigenvalue. However, it tends to become so towards the infinite-volume limit \cite{Sternbeck:2005vs,Sternbeck:2012mf,Cucchieri:2008fc,Cucchieri:2013nja}. It will therefore be assumed that choosing the largest $b$ rather than the smallest eigenvalue will at most increase the finite-volume effects, and will yield the same result in the infinite-volume limit.

\begin{figure}
\includegraphics[width=\linewidth]{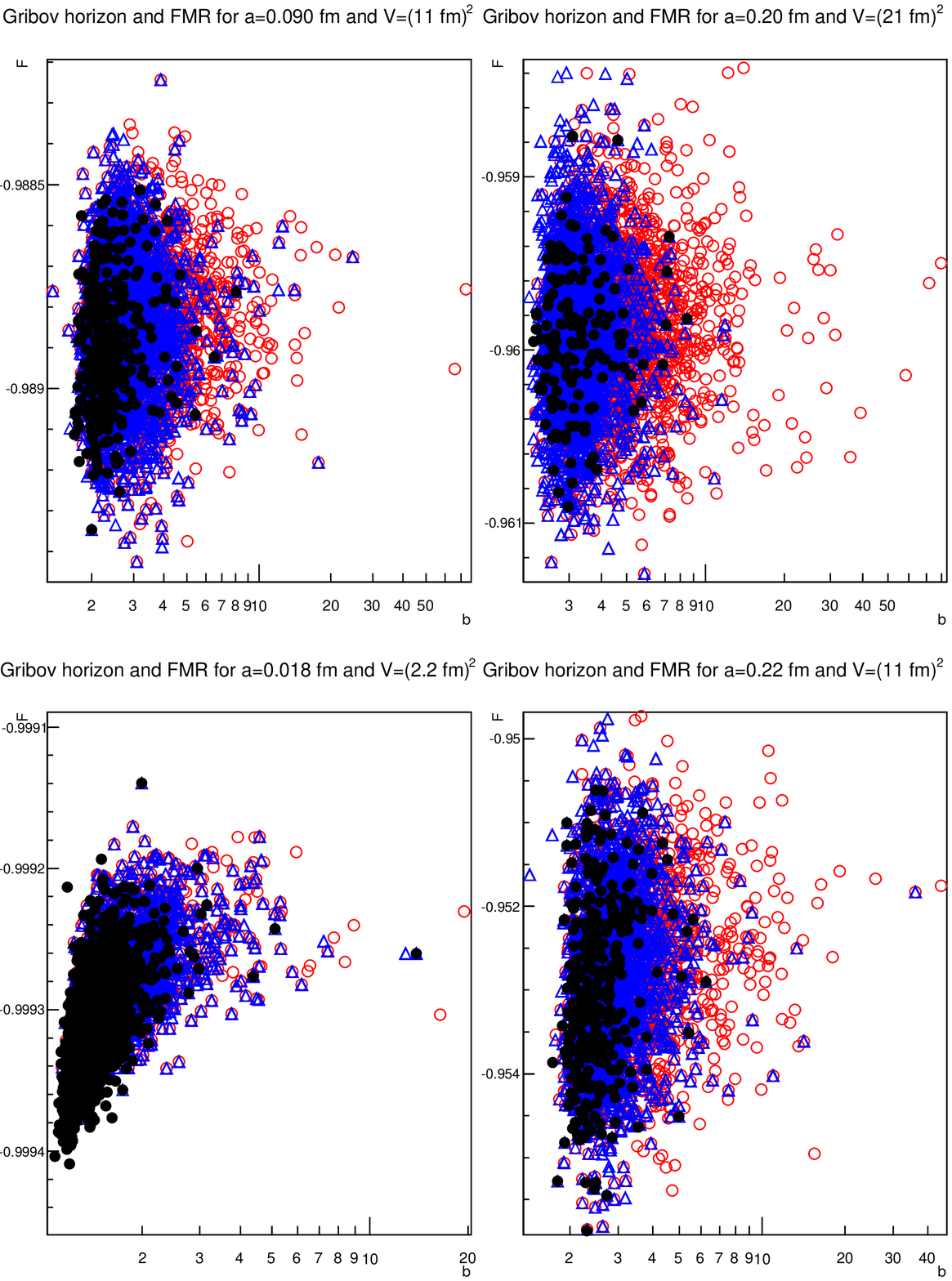}
\caption{\label{fig:hf2} Boundary structure of the Gribov region in two dimensions, based on the found and identified genuine Gribov copies. Open red circles are elements of the horizon, subject to the approximations and assumptions made in the text. Open blue triangles belong to the FMR. Filled black circles possibly belong to both, see text for details. The different panels show results for different discretizations and physical volumes.}
\end{figure}

\begin{figure}
\includegraphics[width=\linewidth]{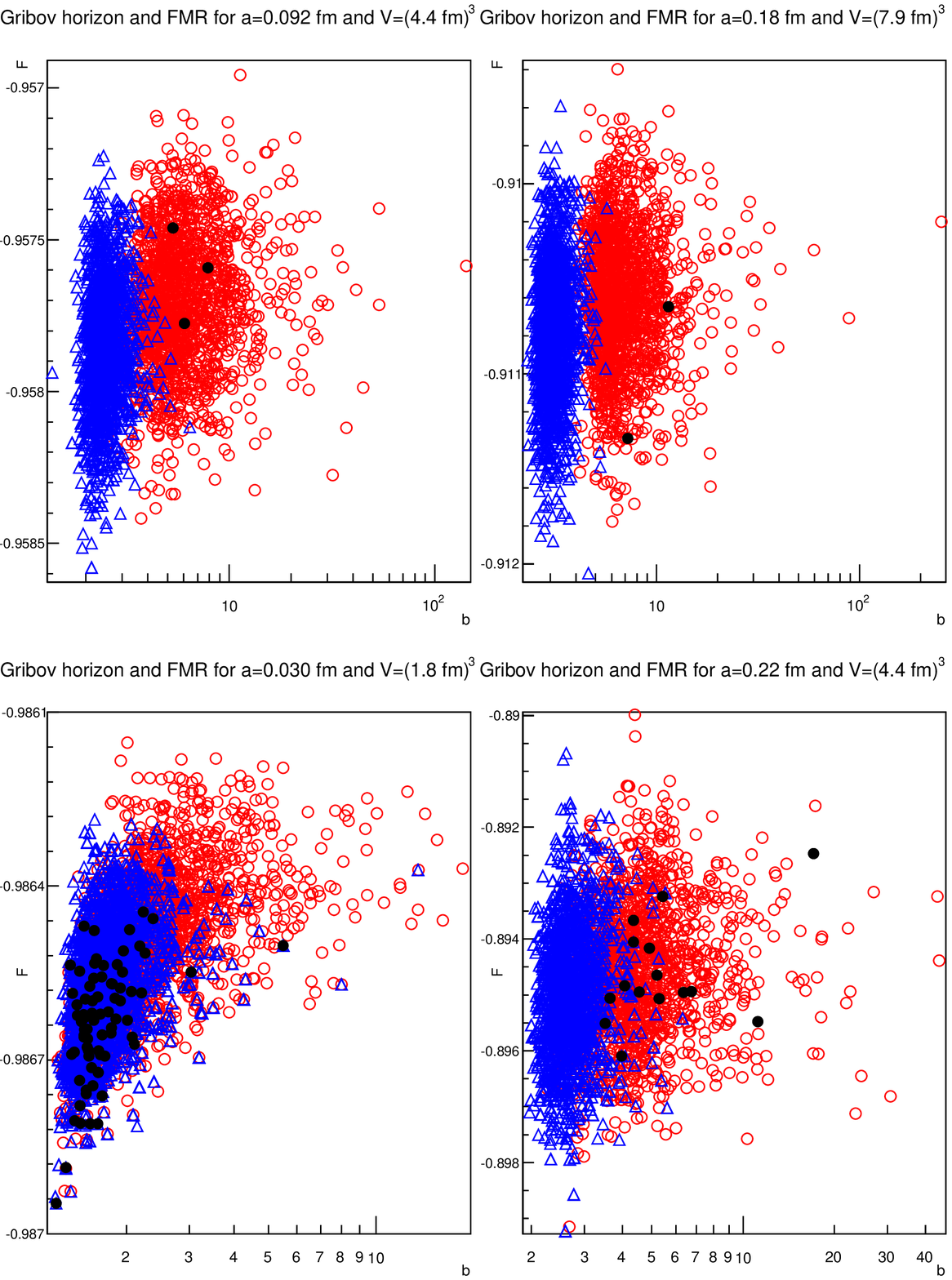}
\caption{\label{fig:hf3} Boundary structure of the Gribov region in three dimensions, based on the found and identified genuine Gribov copies. Open red circles are elements of the horizon, subject to the approximations and assumptions made in the text. Open blue triangles belong to the FMR. Filled black circles possibly belong to both, see text for details. The different panels show results for different discretizations and physical volumes.}
\end{figure}

\begin{figure}
\includegraphics[width=\linewidth]{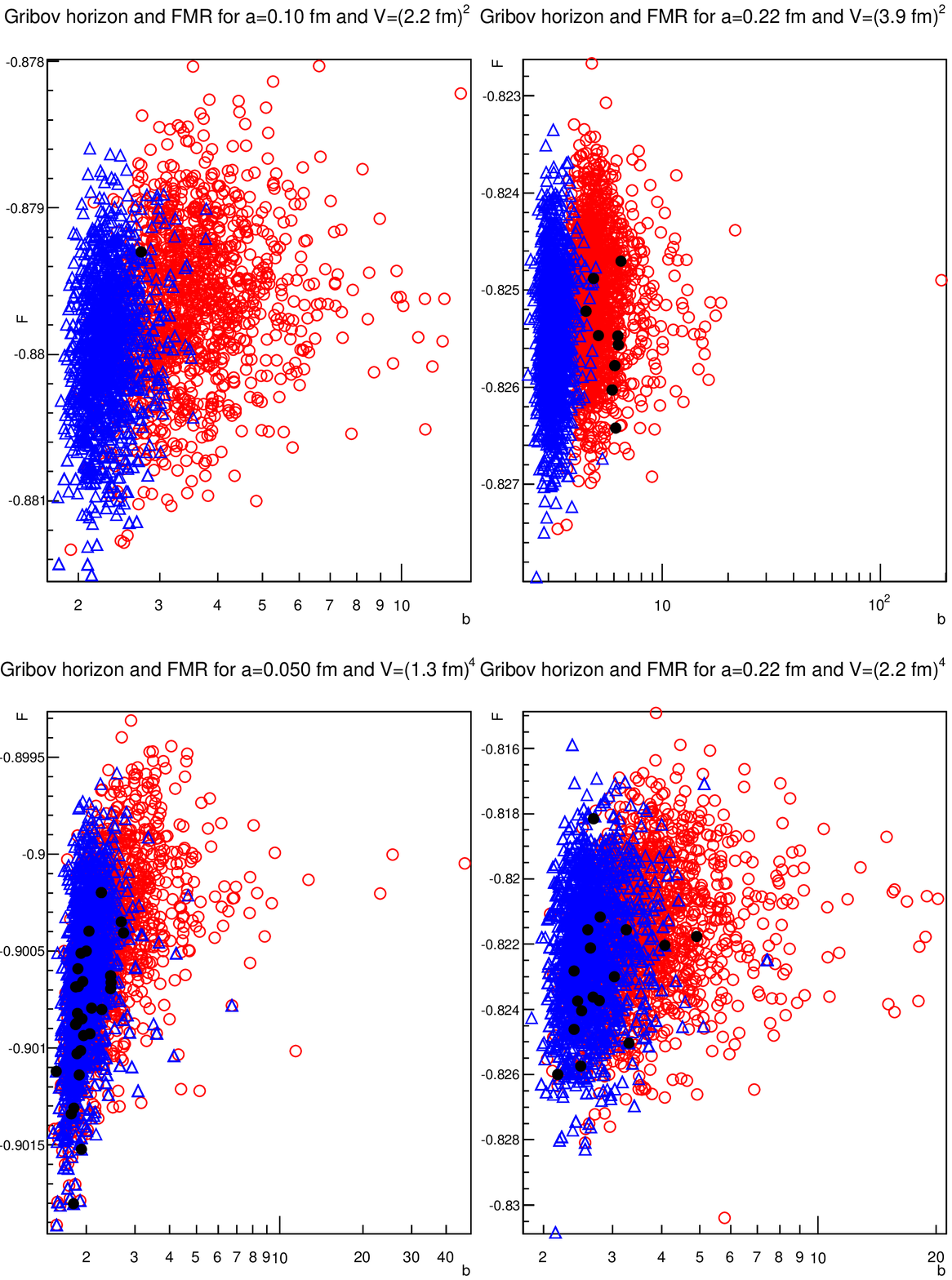}
\caption{\label{fig:hf4} Boundary structure of the Gribov region in four dimensions, based on the found and identified genuine Gribov copies. Open red circles are elements of the horizon, subject to the approximations and assumptions made in the text. Open blue triangles belong to the FMR. Filled black circles possibly belong to both, see text for details. The different panels show results for different discretizations and physical volumes.}
\end{figure}

\afterpage{\clearpage}

The resulting shapes of the boundaries are shown in figures\footnote{First, similar plots for the Gribov horizon can be found in \cite{Sternbeck:2012mf,Maas:2011se}, but without a detailed analysis of other structures, nor with the present systematics.} \ref{fig:hf2}-\ref{fig:hf4}. Of course, given that only a small subset of genuine Gribov copies is found, this can only be a lower limit to the actual horizon, which encloses the displayed horizon.

Once the volume is sufficiently large enough, the rough shape of the horizon in these coordinates is wedge-like. Especially extreme values of $b$ correspond to average values of $F$, while extreme values of $F$ correspond to comparatively small values in $b$. Of the expected non-regular shapes nothing is observed, as expected in a finite-dimensional space. However, as expected after projection, the boundary is such that the interior is connected and the region convex, like the full region itself \cite{Zwanziger:1993dh,Zwanziger:2003cf}. Its finiteness \cite{Zwanziger:1993dh,Zwanziger:2003cf} is evident, but this is trivially so on any finite lattice. There is actually very little difference between the different dimensionalities visible.

It can be expected that there should also be degenerate copies on the horizon, i.\ e.\ more than one Gribov copy of any residual gauge orbit can reside on the boundary. This has been seen even for explicit examples \cite{Maas:2005qt,Sobreiro:2005ec}. It is also found here, i.\ e.\ there are genuine copies with the same value of $b$ but differing value of $F$. The number of such Gribov copies is extremely small, and quickly diminishes with increasing volume, being at the 1\%-level for the largest volumes displayed in dimensions different from two, and at the 10\%-level for two dimensions. However, this is most likely rather an effect from not finding all Gribov copies with increasing volume than that their number actually diminishes. Those found do show a tendency to cluster at comparatively small values of $b$. This is also visible when projecting the Gribov horizon to its coordinates, as is shown in figures \ref{fig:hf2p}-\ref{fig:hf4p} below.

\subsection{The fundamental modular region}

There is another interesting concept in the structure of the first Gribov region, the fundamental modular region \cite{Zwanziger:1982na,Zwanziger:1993dh}. It is defined as the set of all Gribov copies for which $F$ has an absolute minimum on the residual gauge orbit. This region is, in a sense, minimal, as every gauge orbit, which passes through its interior does so at most once \cite{Dell'Antonio:1991xt}, but there may be multiple passages on the boundary for topologically inequivalent copies \cite{vanBaal:1991zw,Dell'Antonio:1991xt}.

Just like the first Gribov region, the FMR is convex, bounded, and contains the origin. Is is therefore completely contained inside the first Gribov region. On a finite volume it is actually expected to be completely inside the interior, while in the infinite-volume limit its boundary and the Gribov horizon should have some overlap \cite{Zwanziger:1993dh,Zwanziger:2003cf}.

The FMR is also shown in figures \ref{fig:hf2}-\ref{fig:hf4}. It exhibits the expected convexity. It is found to be located at small values of $b$, but it is not concentrated at particular values of $F$. The former is somewhat expected, as it should reside in the interior of the Gribov region at finite volume \cite{Zwanziger:1993dh}. However, since the FMR should be located at the absolute minimum of F, it is somewhat surprising to see that it is not concentrated at particularly small values of $F$. This implies that the actual absolute minimum value of $F$ appears to vary strongly between orbits. None of these observations appear to depend strongly on either dimensionality nor discretization. Some effect is seen for different physical volumes, but since the number of Gribov copies so strongly rises it is not possible to make a definite statement whether this is a genuine effect.

Though there is no overlap of the FMR boundary and the horizon in a finite volume, there are points which at the same time have the largest value of $b$ and the smallest value of $F$. However, the number of points found in this way diminishes quickly with volume. Those which are found are located in two dimensions at rather small $b$, but tend to larger $b$ values in higher dimensions and larger volumes. Again, it is not possible to identify whether this is an artifact of too few Gribov copies found or a genuine effect, and should therefore be interpreted, if at all, with great care. The only remarkable observation is that those common points lie at large volumes, in three and four dimensions, at rather large values of $b$ compared to the remainder of the FMR, while this is not the case in two dimensions.

Thus, in total the Gribov horizon and the FMR are found to correspond to the expectations. However, there is a strong correlation between being in the FMR and having a small value of $b$, but no correlation between being either on the horizon or in the FMR with a particular value of $F$. 

Similarly as before, there are Gribov copies with the same (lowest) value of $F$ but differing values of $b$. These should correspond to degenerate copies on the boundary of the FMR. In three and four dimensions, this is again  quickly decreasing with increasing volume, again at the 1\% level at the largest volumes displayed, though much larger on small volumes. Only in two dimensions, this number remains large, and more-or-less volume-independent. These cases show no preference in their distribution on $F$, as is again shown in the projection to the single coordinates below in figures \ref{fig:hf2p}-\ref{fig:hf4p}. 

\subsection{The first Gribov region}

\begin{figure}
\includegraphics[width=\linewidth]{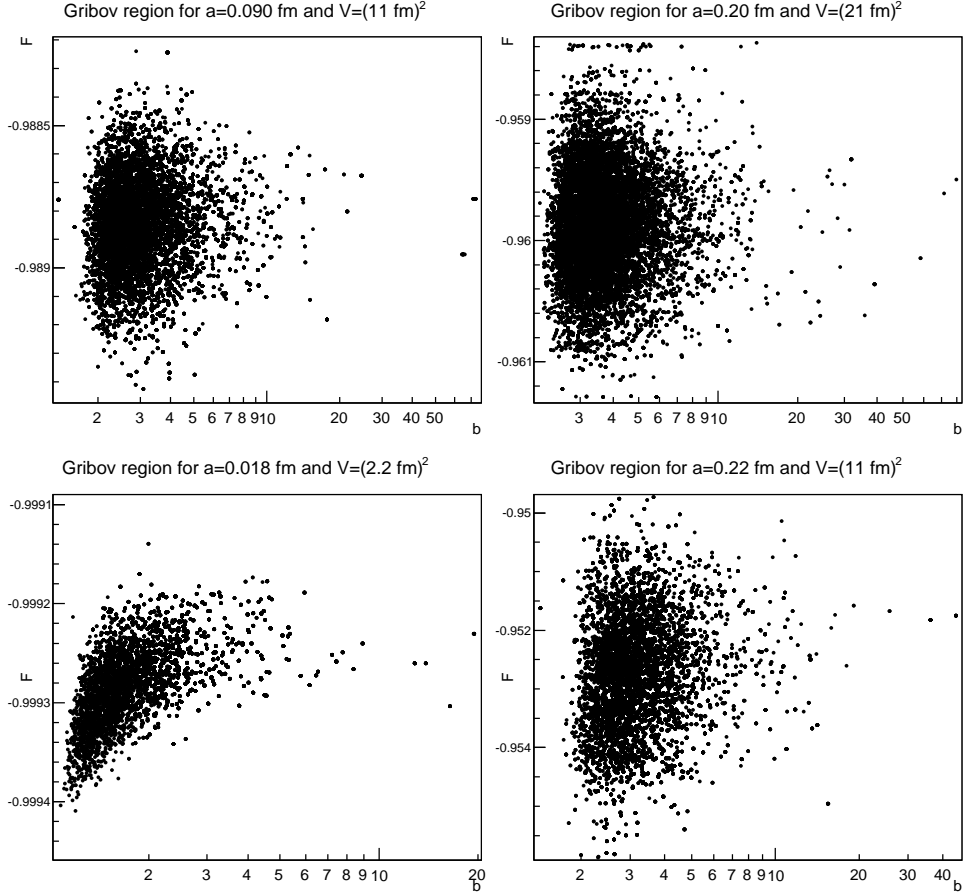}
\caption{\label{fig:g2} The Gribov region in two dimensions, based on the found and identified genuine Gribov copies. Every point is one Gribov copy. The different panels show results for different discretizations and physical volumes.}
\end{figure}

\begin{figure}
\includegraphics[width=\linewidth]{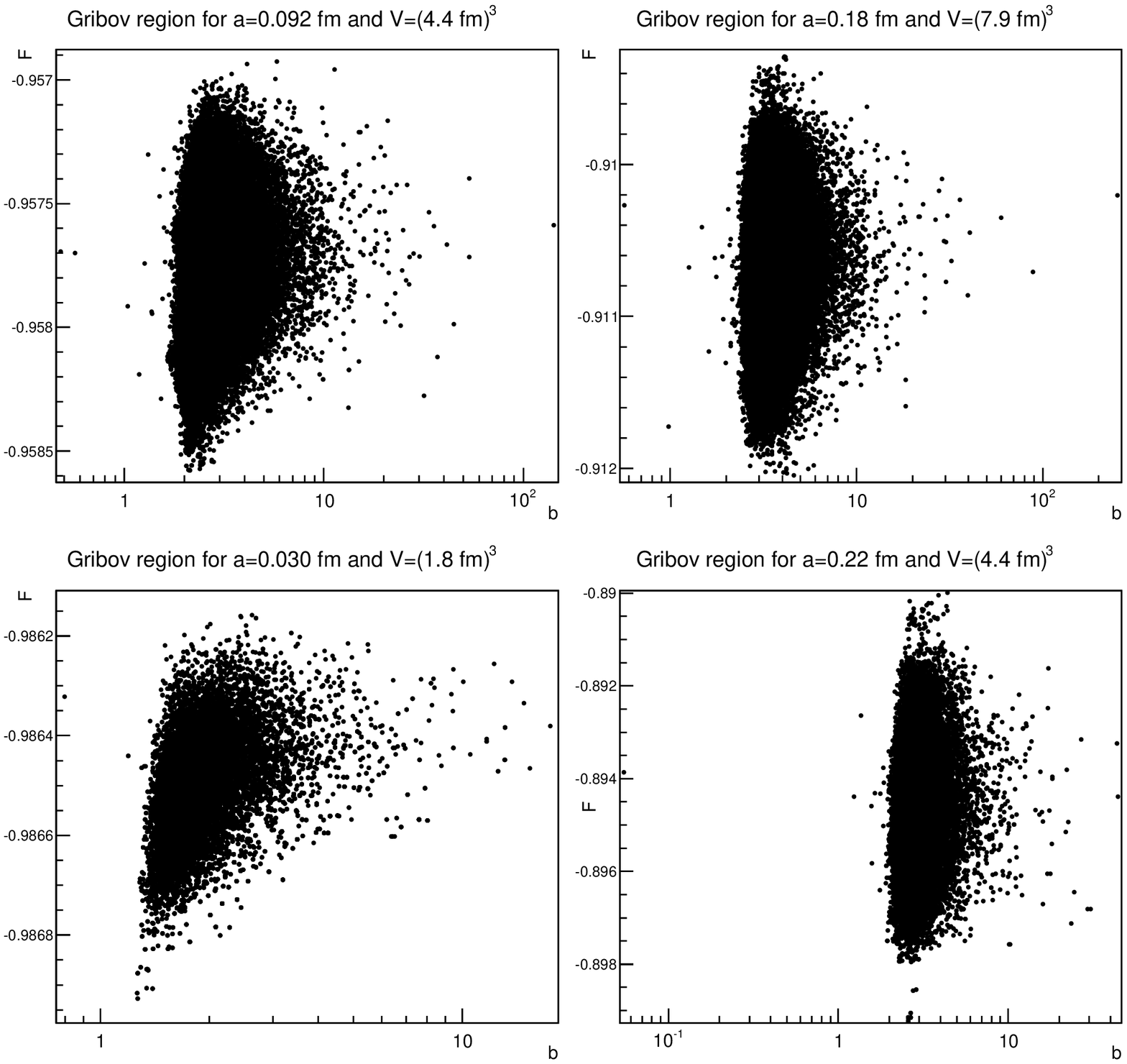}
\caption{\label{fig:g3} The Gribov region in three dimensions, based on the found and identified genuine Gribov copies. Every point is one Gribov copy. The different panels show results for different discretizations and physical volumes.}
\end{figure}

\begin{figure}
\includegraphics[width=\linewidth]{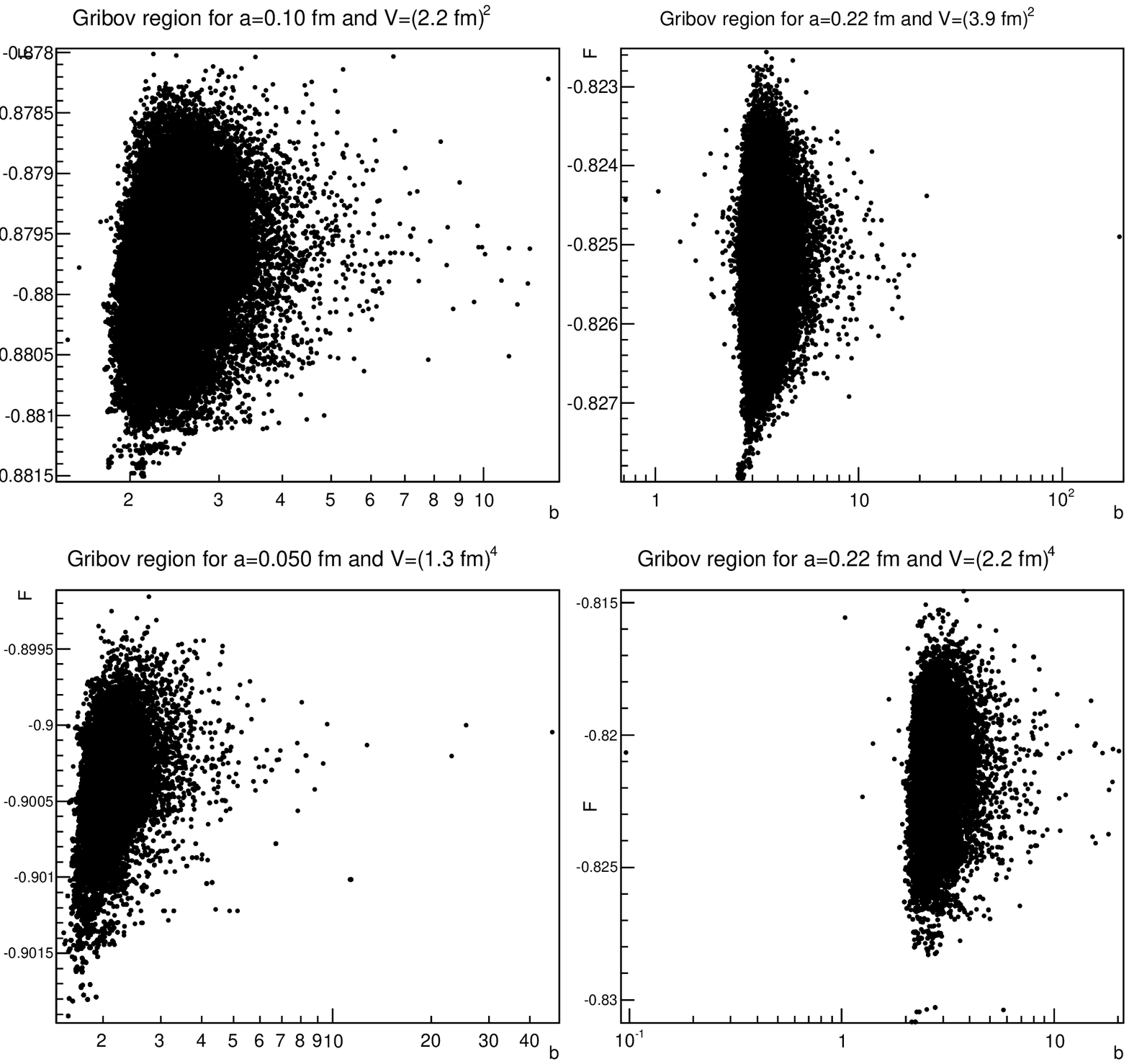}
\caption{\label{fig:g4} The Gribov region in four dimensions, based on the found and identified genuine Gribov copies. Every point is one Gribov copy. The different panels show results for different discretizations and physical volumes.}
\end{figure}

A final view is now on the first Gribov region as a whole, in the same manner as in the previous section. Every genuine Gribov copy found is displayed in the figures \ref{fig:g2}-\ref{fig:g4}. Except for the appearance of many more copies at smaller values of $b$, as expected, the shape remains essentially the same, a wedge, especially the convexity is still very well visible. The boundaries are still somewhat frayed, probably a result of both finite statistics and a finite number of Gribov copies per orbit.

Interestingly, the full regions sweep out the same range of $F$ values as the FMR. Thus, the variation of $F$ inside the FMR and in the full region is actually rather similar.

\begin{figure}
\includegraphics[width=\linewidth]{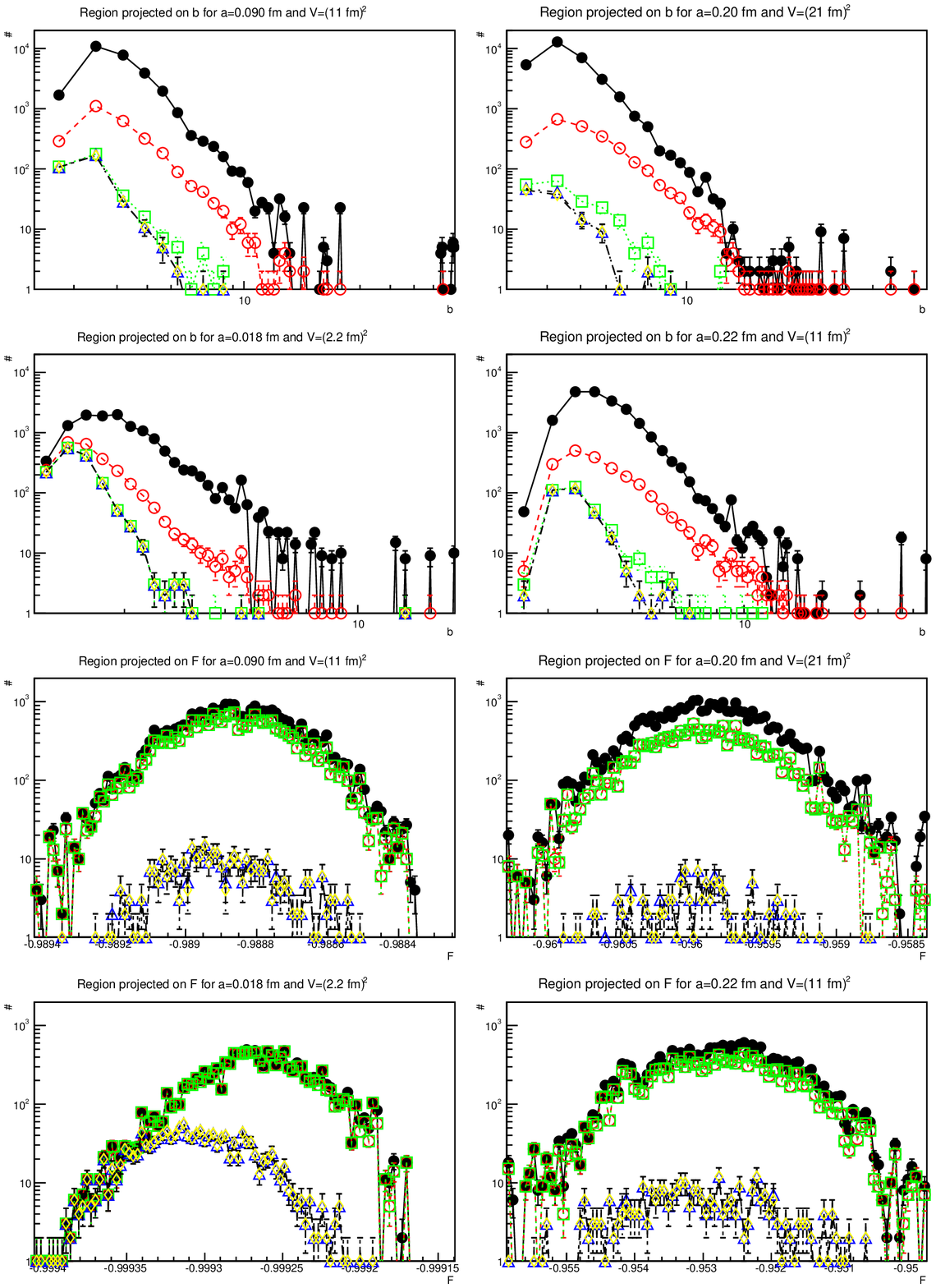}
\caption{\label{fig:hf2p} Projections of the Gribov region to its coordinates $b$ (top two rows) and $F$ (bottom two rows) for various lattice spacings in two dimensions. Black dots are the full region, open red dots indicate the Gribov horizon and green open squares are degenerate points on the Gribov horizon in the projection on $b$, while the same symbols denote the FMR and degenerate points on the FMR in the projection on $F$. Open blue triangles are element of the common boundary in both projections, and open yellow triangles are degenerate common points.}
\end{figure}

\begin{figure}
\includegraphics[width=\linewidth]{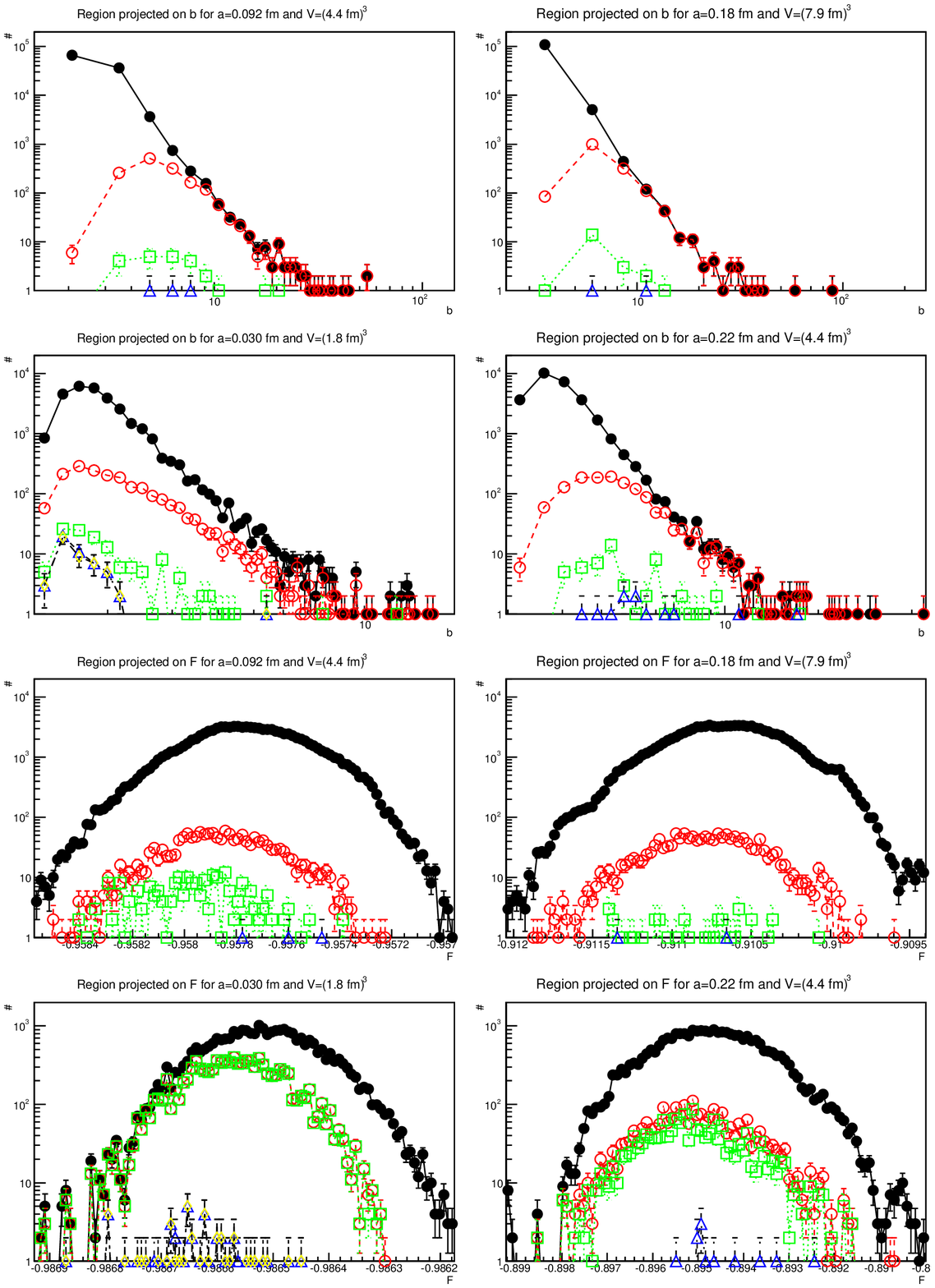}
\caption{\label{fig:hf3p}Projections of the Gribov region to its coordinates $b$ (top two rows) and $F$ (bottom two rows) for various lattice spacings in three dimensions. Black dots are the full region, open red dots indicate the Gribov horizon and green open squares are degenerate points on the Gribov horizon in the projection on $b$, while the same symbols denote the FMR and degenerate points on the FMR in the projection on $F$. Open blue triangles are element of the common boundary in both projections, and open yellow triangles are degenerate common points.}
\end{figure}

\begin{figure}
\includegraphics[width=\linewidth]{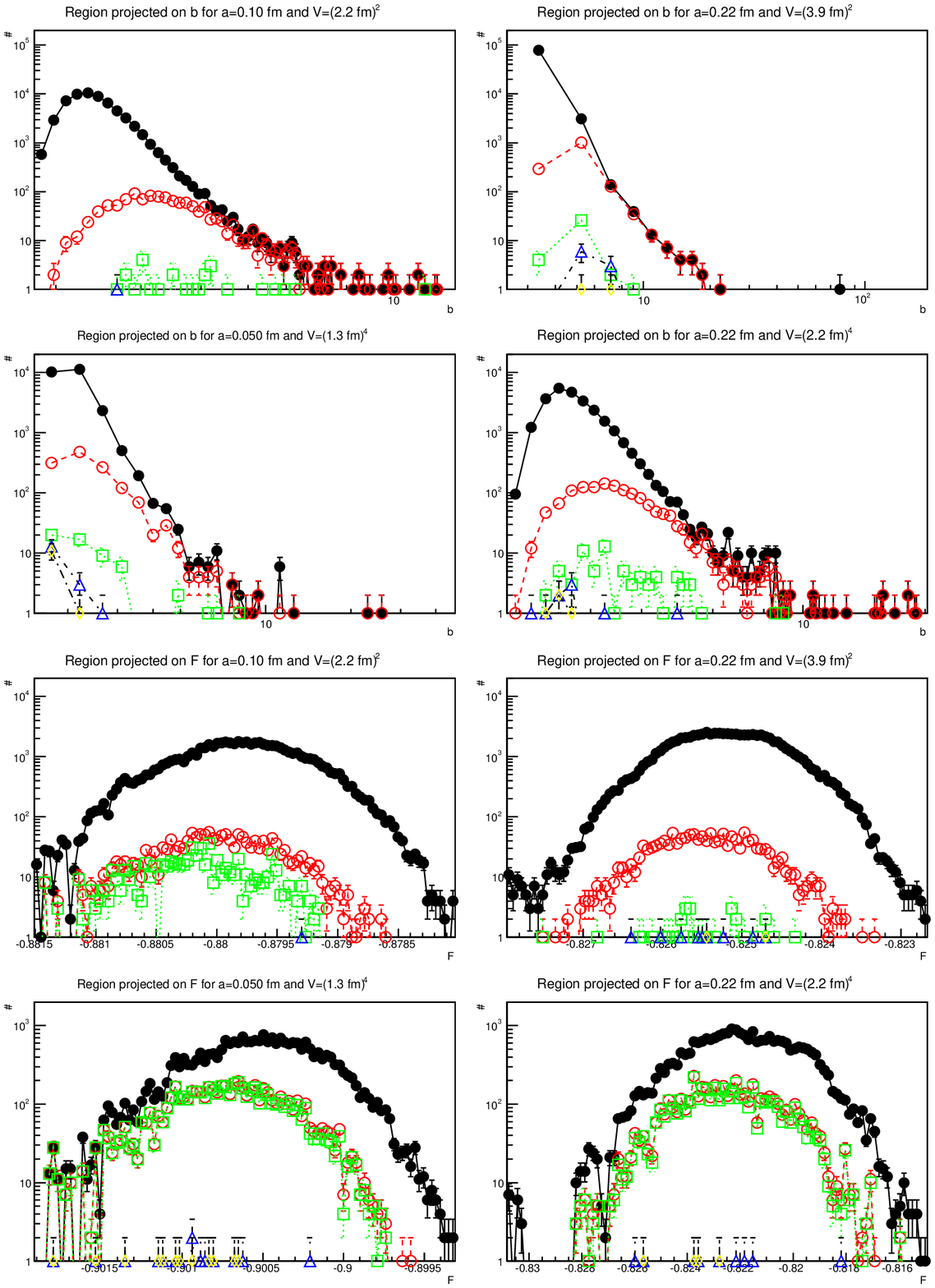}
\caption{\label{fig:hf4p}Projections of the Gribov region to its coordinates $b$ (top two rows) and $F$ (bottom two rows) for various lattice spacings in four dimensions. Black dots are the full region, open red dots indicate the Gribov horizon and green open squares are degenerate points on the Gribov horizon in the projection on $b$, while the same symbols denote the FMR and degenerate points on the FMR in the projection on $F$. Open blue triangles are element of the common boundary in both projections, and open yellow triangles are degenerate common points.}
\end{figure}

All of these results are also visible in the projection of the first Gribov region to the two axes, completing the tomography, and shown in figures \ref{fig:hf2p}-\ref{fig:hf4p}. There, it is explicitly seen how the horizon saturates the large-$b$-value region, but there are still many orbits where it has rather low values. Also, degenerate points are located at somewhat lower values of $b$ than non-degenerate ones. At the same time, the FMR is rather evenly distributed. Interestingly, in two dimensions almost all FMR points are degenerate boundary points, and there is almost no interior. That is very different in dimensions greater than two, where the degenerate boundary points are much less and seem to be mainly located at smaller values of $F$. The common boundary, including degenerate points, is also shown, and is mainly located at slightly smaller values of both $b$ and $F$ than the non-common boundary points.

\section{Summary}\label{s:sum}

The present study is the first systematic investigation of the (Landau-gauge) first Gribov region using a two-dimensional projection, following up the first investigations in \cite{Maas:2011se,Sternbeck:2012mf}. It also demonstrates explicitly that the number of Gribov copies is very large, even on moderately to small lattice systems, so that any investigation of Gribov copy effects can only be considered as statements about lower limits, and extrapolation is necessary.

The explicit tomography of the first Gribov region, including its most prominent features of the Gribov horizon and the FMR and its boundary, can hence only be considered to be something of a sketch. However, this sketch already reveals many of the properties expected, like the convexity. It also shows some rather surprising features, especially that the FMR has no distinguished tendency to be at small values of $F$, but has even on the largest volumes a strong preference for small $b$ values. Also, with all due caveats, the common boundary between FMR and the Gribov region seems to have a preference for smaller values of both $b$ and $F$.

This knowledge is a very suitable starting point to develop various extended gauge conditions, aimed at emphasizing different properties of gauge-dependent correlation functions along the lines of \cite{Maas:2011se,Maas:2013vd,Mehta:2009zv,Neuberger:1986xz,vonSmekal:2008en,vonSmekal:2008es,vonSmekal:2007ns,Serreau:2012cg,Parrinello:1990pm,Serreau:2013ila}, essentially gauge engineering. This permits, just as in perturbation theory, to tailor the features of gauge-dependent correlation function to  being technically best suited for further calculations, e.\ g.\ using functional methods \cite{Maas:2011se}.\\

\no{\bf Acknowledgments}\\

I am grateful to M.\ Huber for a critical reading of the manuscript and helpful remarks. This work was supported by the DFG under grant numbers MA 3935/5-1, MA-3935/8-1 (Heisenberg program) and the FWF under grant number M1099-N16. Simulations were performed on the HPC cluster at the Universities of Jena and Graz. The author is grateful to the HPC teams for the very good performance of the clusters. The ROOT framework \cite{Brun:1997pa} has been used in this project.

\bibliographystyle{bibstyle}
\bibliography{bib}


\end{document}